\documentclass[prb,reprint,aps,twocolumn]{revtex4-1}
\usepackage[T1]{fontenc}
\usepackage[utf8]{inputenc}
\usepackage[english]{babel}
\usepackage{float}
\usepackage{tabularx}
\usepackage{caption}
\usepackage{subcaption}
\usepackage{graphicx}
\graphicspath{{png/}}
\usepackage{color}
\usepackage{amsmath}
\usepackage{amssymb,amsfonts,textcomp}
\usepackage{hyperref}
\usepackage{booktabs}
\usepackage[outdir=./]{epstopdf}
\usepackage{subcaption}
\usepackage{amsmath}
\usepackage{soul}
\usepackage{placeins}

\hypersetup{colorlinks = true, citecolor = blue, breaklinks = true}

\usepackage[resetlabels]{multibib}
\newcites{SI}{References}

\begin{document}

\title{Density functional theory study of the oxygen evolution activity on Sr$_2$TaO$_3$N surfaces}
\date{\today}
\author{Maria Bouri}
\affiliation{Department of Chemistry and Biochemistry, University of Bern, Freiestrasse 3, CH-3012 Bern, Switzerland}
\author{Ulrich Aschauer}
\affiliation{Department of Chemistry and Biochemistry, University of Bern, Freiestrasse 3, CH-3012 Bern, Switzerland}

\begin{abstract}
Solar water splitting has attracted much attention as a clean and renewable route to produce hydrogen fuel. Since the oxygen evolution half-reaction (OER) requires high overpotentials, much research has focused on finding catalyst materials that minimize this energy loss. Oxynitrides with a layered perovskite structure have the potential to combine the superior photocatalytic properties of layered perovskite oxides with enhanced visible-light absorption caused by the band gap narrowing due to less electronegative nitrogen ions. In this paper, we study the OER on the (001) and (100) surfaces of the layered oxynitride Sr$_2$TaO$_3$N using density functional theory (DFT) calculations to obtain the OER free energy profiles and to determine the required overpotentials at various sites on each surface. We find that the reconstructed grooved (100) surface is most relevant for photocatalysis due to suitable band-edge positions combined with a low overpotential and good carrier mobility perpendicular to the surface.
\end{abstract}

\maketitle

\section*{Introduction}
Over recent decades, an ever increasing energy demand together with the negative environmental impact of existing fossil and nuclear energy sources has lead to an intense search for clean and renewable forms of energy and ways to store it. Water splitting to produce hydrogen (H$_2$) fuel with the energy stored in the H-H chemical bond has attracted much attention in this domain \cite{turner2004sustainable}. Fuel cells can convert the produced H$_2$ back into electricity for various applications with just water as the byproduct. When the energy required for water splitting is provided by solar radiation, the procedure is sustainable and the produced H$_2$ is environmentally neutral. 

The overall water-splitting reaction consists of two half-reactions, the H$_2$ evolution reaction (HER) being driven by excess electrons and the O$_2$ evolution reaction (OER) by excess holes. In photocatalysis, recent studies have focused on devices in which a semiconductor photoanode catalyzes the OER and a metal cathode the HER \cite{Gratzel:79530}. The excess carriers are generated on the photoanode by cross-gap electron excitation due to incident photons, the excitation providing a potential corresponding to the band gap of the absorbing semiconductor. Ideally an OER photocatalyst will have to provide a potential of 1.23 V but in reality larger potentials are required, the difference being known as the overpotential. Besides interface losses, the overpotential stems from the fact that due to material-dependent adsorbate binding energies the individual reaction steps of the four-step OER have free-energy differences that deviate from the ideal value of 1.23 eV \cite{Rossmeisl2005,Valdes2008}. Scaling relations between the different adsorbate adsorption energies lead to minimum theoretical overpotentials of 0.37 V \cite{koper2013theory,man2011universality}, making this half-reaction the bottleneck for water splitting \cite{greeley2012road, gimenez2016photoelectrochemical}. Rendering this technology economically viable crucially requires photoanodes that simultaneously absorb visible light (band gaps below 3 eV) and provide the required overpotential (band gaps above 1.7 eV and adequate band edge positions) while having overpotentials as close as possible to the theoretical minimum. 

\begin{figure}
	\centering
	\includegraphics[width=0.9\columnwidth]{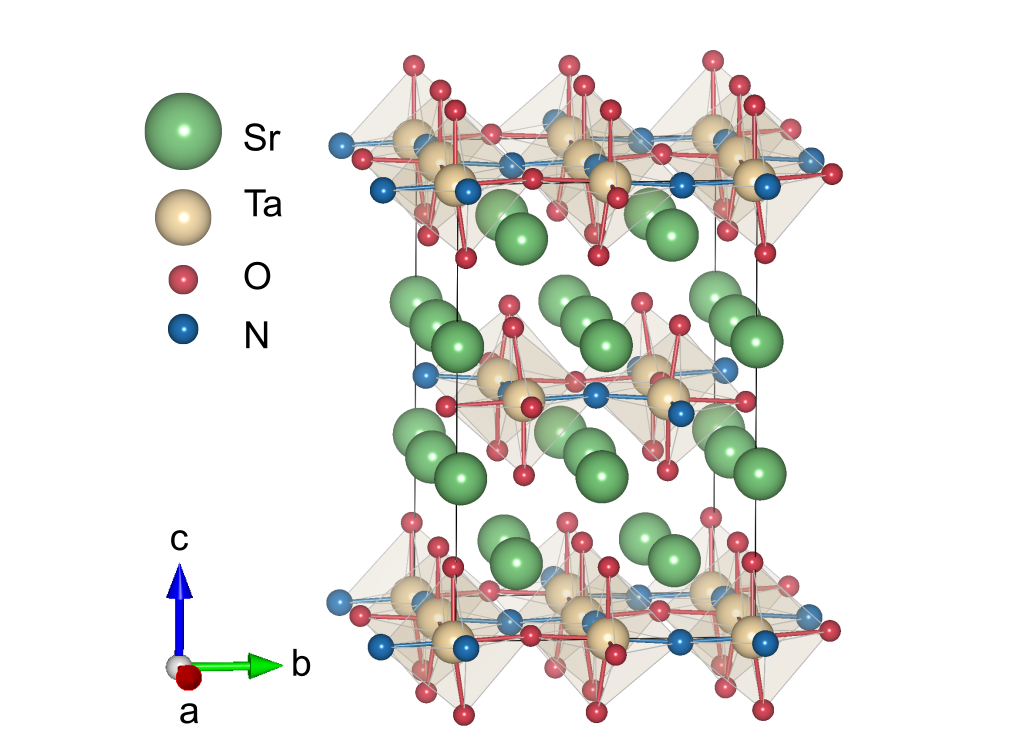}
	\caption{Energetically favoured configuration of the 56-atom bulk Sr$_2$TaO$_3$N unit cell with octahedral rotations and the nitrogen atoms adopting a \textit{cis} order on the equatorial sites.}
	\label{fig:bulk}
\end{figure}

Perovskite structured oxides with d$^0$-transition metal B site cations were shown to be efficient photocatalysts \cite{Kudo2009, Takata2015}, but their wide band gaps make them inappropriate for visible-light absorption. Substitution of oxygen by less electronegative nitrogen leads to N 2\textit{p} states above the O 2\textit{p} states at the top of the valence band (VB). Since the conduction band (CB) states remain nearly unaffected by this substitution \cite{Fuertes2015a, Porter2014, ji2005photocatalytic}, the band gap is reduced and these so-called oxynitrides are suitable for visible-light absorption. 

Perovskite materials can occur in multiple layered structure types, among them Ruddlesden-Popper (RP) \cite{ruddlesden1957}, Dion-Jacobson \cite{dion1981nouvelles,jacobson1985interlayer}, Aurivillius phases \cite{aurivillius1953structure} and the Lichtenberg phases \cite{lichtenberg2008synthesis, valdez2019origin}. Some of these layered structure types were shown to have higher photocatalytic H$_2$ and O$_2$ production compared to chemically similar non-layered perovskite oxides \cite{Kudo2009}. A computational screening of layered perovskite oxides and oxynitrides in the RP structure indeed showed that these materials have appropriate band gaps and that their VB and CB edges straddle the water redox levels \cite{Castelli2013a}. Beside these bulk criteria the surface structure as well as the oxidising adsorbates that form under application conditions are expected to affect the OER pathways and the resulting activity of the catalyst. Up to date, computationally studies of the OER mechanism have mostly focused on metal oxide surfaces \cite{man2011universality, Valdes2008, bajdich2013theoretical}, while only few studies investigate oxynitride surfaces \cite{montoya2015theoretical, ouhbi2018water} and no reports exist for surfaces of layered RP oxynitrides.

Here we use density functional theory (DFT) calculations to determine the thermodynamically most stable terminations of the (001) and (100) surfaces of the RP oxynitride Sr$_2$TaO$_3$N (bulk structure in Fig. \ref{fig:bulk}) under photochemical conditions. We then determine the overpotentials of different OER mechanisms via Gibbs free-energy differences of the individual reaction steps on the most relevant terminations. Our findings show that a reconstructed (100) surface is most relevant for photocatalysis due to its stability, a low overpotential when covered with oxygen adsorbates and a valence band maximum that provide sufficient potential to drive the OER. Moreover, the reconstructed RP (100) surface shows better OER activity under operating conditions than the (001) surface of a non-layered material based on the same elements.

\section*{Computational Approach}

All density functional theory (DFT) calculations were carried out with the Quantum ESPRESSO package \cite{Giannozzi2009} using the Perdew-Burke-Ernzerhof (PBE) \cite{Perdew1996} exchange-correlation functional. Kinetic-energy cut-offs of 35 Ry for the plane wave basis and 280 Ry for the augmented density were used for all calculations. Ultrasoft pseudopotentials \cite{Vanderbilt1990} with Sr (4s, 4p, 4d, 5s, 5p), Ta (5s, 5p, 5d, 6s, 6p), O (2s, 2p), N (2s, 2p) and H (1s) as valence electrons were used to describe electron-nuclear interactions. Surface geometries were optimised with a force convergence threshold of 0.025 eV/\AA. Reciprocal space was sampled by $4\times4\times1$ and $4\times2\times1$ Monkhorst-Pack\cite{Pack1977} \textit{k}-point meshes for the (001) and (100) surfaces respectively that have lateral dimensions of $8.188\times 8.188$ \AA\ and $8.194\times 12.526$ \AA. 

\begin{figure}
	\centering
	\includegraphics[width=0.9\columnwidth]{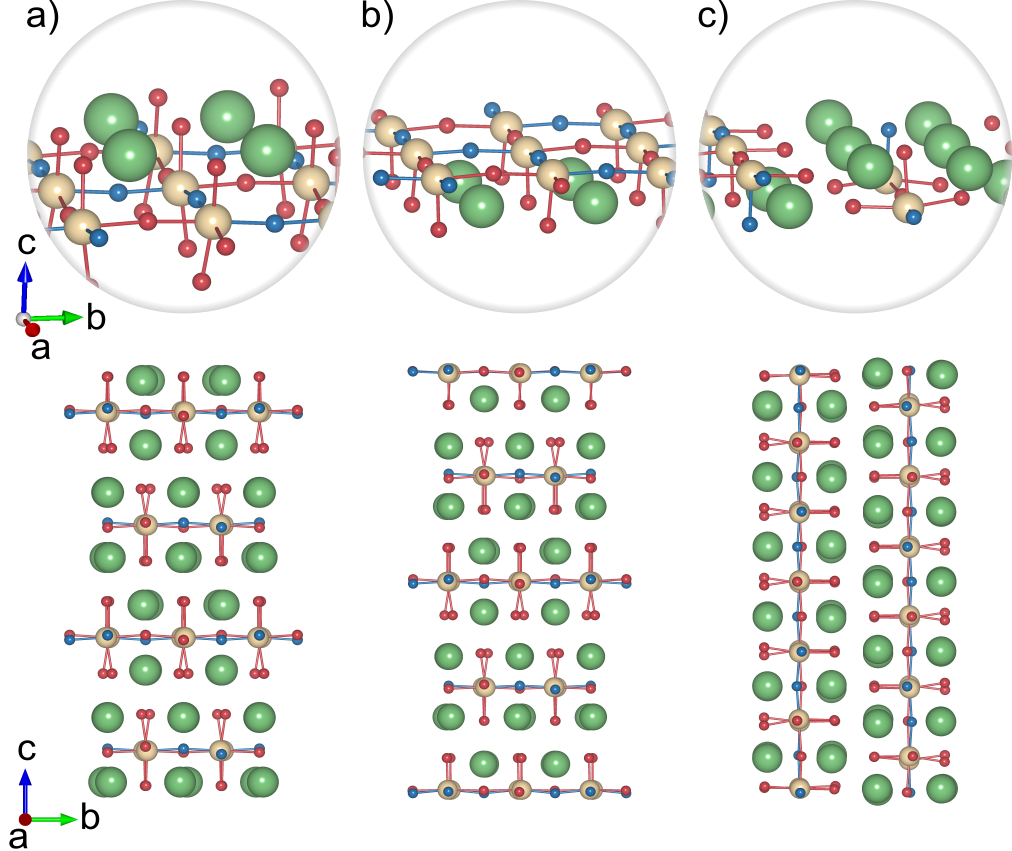}
	\caption{Side and perspective view of a) the SrO-terminated (001), b) the TaON-terminated (001) and c) the (100) surfaces of Sr$_2$TaO$_3$N.}
	\label{fig:surfaces}
\end{figure}

Surfaces were constructed by cleaving the fully optimised bulk structure (Fig. \ref{fig:bulk}) with the lowest energy \textit{cis} nitrogen order and octahedral-rotation distortions \cite{bouri2018bulk}. The SrO-terminated (001) slab (Fig. \ref{fig:surfaces} a) is constructed by cleaving perpendicular to the [001] direction between two consecutive SrO layers. The resulting slab is symmetric and stoichiometric with 12 atomic layers and a thickness of 22.6 \AA. During relaxation the two middle SrO layers are kept fixed. The surface exposes four undercoordinated Sr atoms, surrounded by O atoms, that represent active sites for the OER. The TaON-terminated (001) slab (Fig. \ref{fig:surfaces} b) is created cleaving perpendicular to the [001] direction between a SrO and a TaON layer and adding one extra TaON-layer at the bottom of the slab, resulting in a non-stoichiometric but symmetric 13-layer slab with a thickness of 24.8 \AA. The middle three atomic layers are kept fixed during relaxation. The surface exposes four undercoordinated Ta atoms, surrounded by O and N atoms, on which the OER can take place. The N atoms form a \textit{cis} order in the \textit{ab} plane. The (100) slab (Fig. \ref{fig:surfaces} c) is obtained by cleaving the bulk perpendicular to the [100] direction resulting in a 24.4 \AA\ thick, symmetric and stoichiometric slab with 13 atomic layers. Here, the three middle layers are kept fixed during structural optimization. In this case, the surface exposes four undercoordinated Sr atoms and two undercoordinated Ta atoms that can act as reaction sites for the OER. A vacuum of 10 \AA\ is added to all slabs to separate the periodic images along the surface normal direction. A dipole correction \cite{bengtsson1999dipole} was also used along the surface normal direction. Surface Pourbaix diagrams and OER free-energy profiles were computed using the established approach by N{\o}rskov \cite{Valdes2008} (see supporting information section \ref{sec:SI_methods}). We note that, as discussed in the supporting information, the theoretical overpotential is independent of the pH and for simplicity, we therefore consider pH=0 throughout this work.

\section*{Results and discussion}

\subsection*{TaON-terminated (001) surface}

\begin{figure}
	\centering
	\includegraphics[width=0.9\columnwidth]{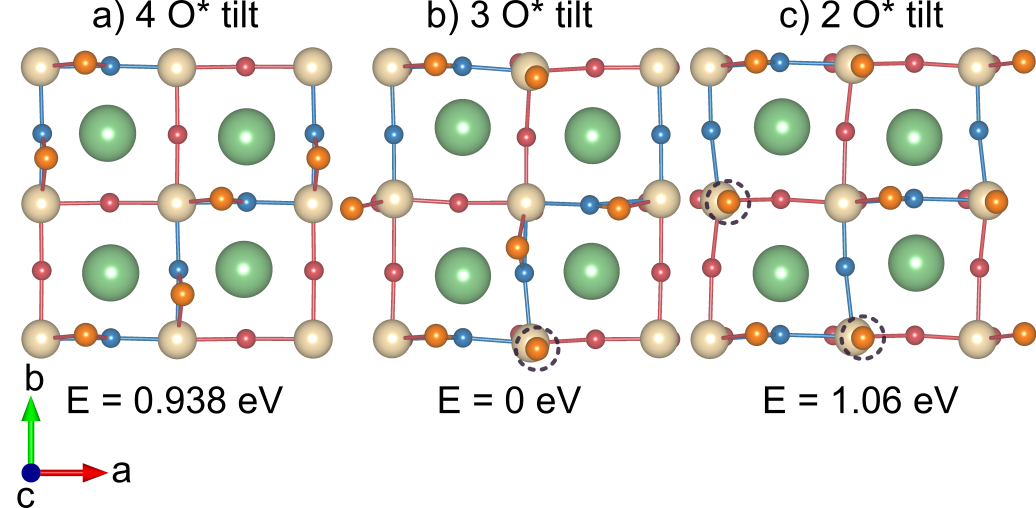}
 	\caption{Top views of different configurations of the fully O*-covered TaON-terminated (001) surface: a) four O* atoms tilted towards surface N atoms following the \textit{zig-zag} N-anion order, b) three O* atoms tilted and one O* upright and c) two O* atoms tilted and two O* atoms upright (diagonally aligned). Orange spheres denote adsorbate O* atoms and dotted circles indicate the upright O* atoms.}
	\label{fig:TaON_full_figures}
\end{figure}

Starting from the clean TaON-terminated (001) slab, we investigate the free-energy changes associated with different coverages of hydroxyl and oxygen adsorbates. While for surfaces covered with OH*, we observe the adsorbates to remain nearly upright or to lean slightly towards neighboring N atoms, we find for the fully O*-covered surface a more complex surface adsorbate structure. At high coverage, O* adsorbates tend to tilt significantly to bond with surface N atoms. We consider the situations where all four O* are tilted (4 O* tilt), three O* are tilted and one remains upright (3 O* tilt) and two of them tilt with two remaining upright (2 O* tilt) as shown in Fig. \ref{fig:TaON_full_figures}. Somewhat surprisingly, we find that the 3 O* tilt configuration is more stable than the 4 O* tilt structure by 0.93 eV, while the 2 O* tilt configuration is significantly less stable. By comparing the density of states (DOS) of the 3 O* tilt and 4 O* tilt configurations (see supplementary Fig. \ref{fig:pdos_O*_taon_surf_layer}), we see that while the latter has equivalent states associated with all adsorbates, adsorbate states in the former become highly unequal and we see the appearance of an unoccupied N-O state in the band gap just above the Fermi energy. Simultaneously, we see a very marked state associated with the upright O* appearing at lower energies than the states originating from the other O*. This can be interpreted as a charge transfer from $\pi^*$ antibonding N-O states (see supplementary Fig. \ref{fig:ildos_vb_peaks}) to lower lying states associated with the upright O*, resulting in stronger N-O bonds and hence a lower total energy. There is hence a balance between the number of N-O bonds and associated bonding and antibonding states and upright O* states, which is optimal for the 3 O* tilt case.

\begin{figure} 
	\centering
	\includegraphics[width=0.9\columnwidth]{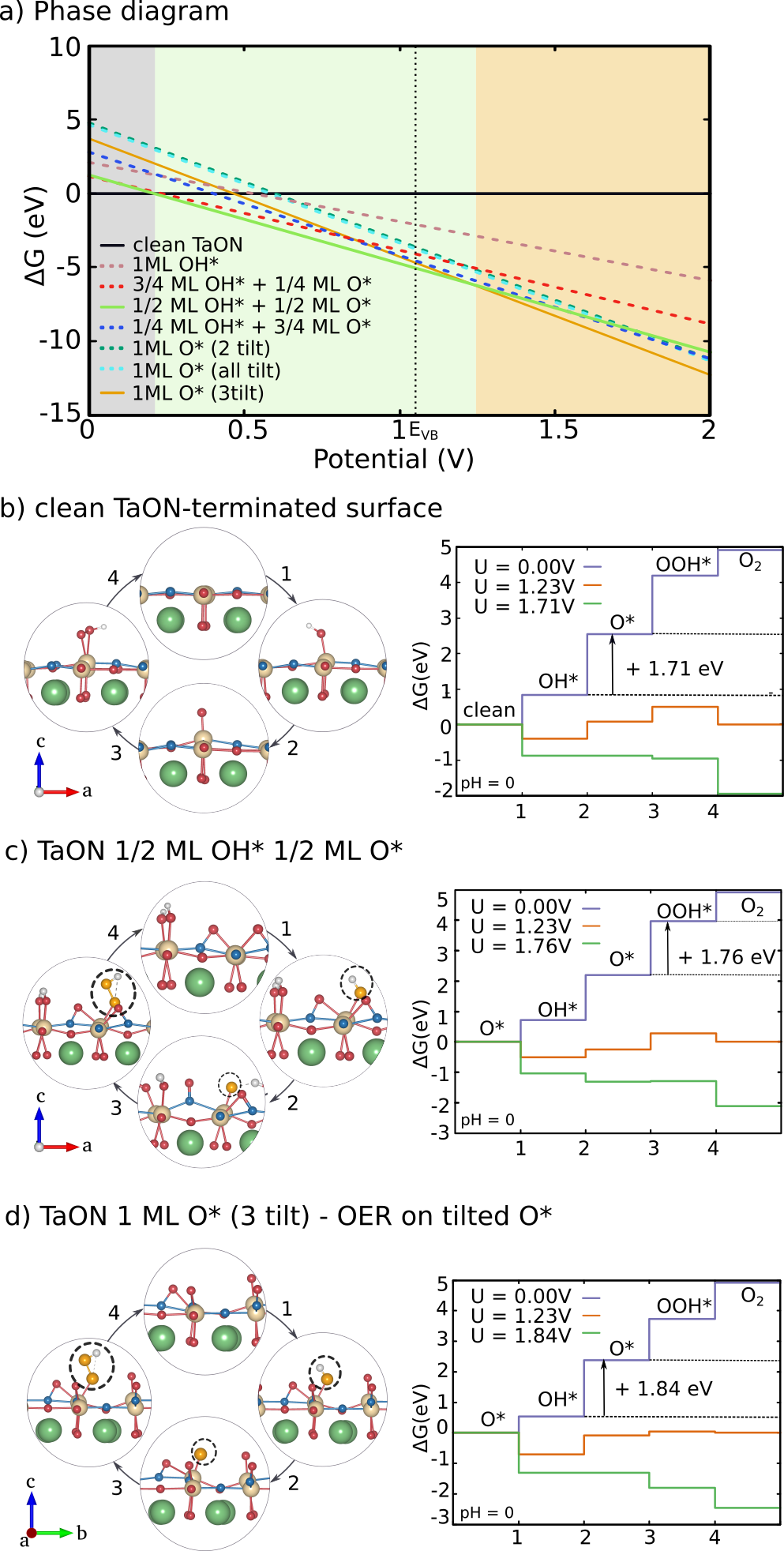}
	\caption{a) Surface Pourbaix diagram of the TaON-terminated (001) surface as well as OER steps (left) and Gibbs free energy diagrams (right) for b) the clean surface, c) the surface covered with 1/2 ML OH* and 1/2 ML O* and d) with 1 ML O* considering the tilted O* as reaction site.}
	\label{fig:TaON_panel}
\end{figure}

In Fig. \ref{fig:TaON_panel} a) we report the energies of surfaces with different adsorbate coverages and configurations with respect to the clean surface as a function of the potential and at pH=0. At potentials close to zero the clean (no adsorbates) surface is most stable, while we observe terminations with a higher coverage of more oxidising adsorbates (O* rather than OH*) to become increasingly more stable at higher potentials. In particular, our calculations predict that for potentials above 0.2 V the surface is covered with 1/2 monolayer (ML) OH* and 1/2 ML O* where we also observe the O* atoms to tilt and bond with surface N atoms (see supplementary Fig. \ref{fig:2O_2OH_taon}). For potentials above 1.2 V the surface is covered with 1 ML of O*, assuming the 3 O* tilt structure (Fig. \ref{fig:TaON_full_figures} b) discussed above.

We first investigate the OER on the clean TaON-terminated (001) surface that is relevant at potentials below 0.2 V and calculate the free-energy differences of the four individual proton-coupled-electron transfer (PCET) steps. The surface configurations during the OER are shown in Fig. \ref{fig:TaON_panel} b. Step 1 corresponds to the deprotonation of one H$_2$O molecule, the resulting OH fragment remaining vertically adsorbed on the Ta site and leading to a small outwards relaxation. In step 2, the OH* is further deprotonated to O*, which remains vertically adsorbed on top of the surface Ta atom, while leading to a more marked outward relaxation. The deprotonation of another water molecule in step 3 and the association of the resulting OH fragment with the adsorbed O* leads to the formation of an OOH* on the Ta site which, in step 4 after a final deprotonation, desorbs as an O$_2$ molecule leaving the surface clean from adsorbates. The free energy diagram (Fig. \ref{fig:TaON_panel} b) reveals that the overpotential-determining step (ODS) is the formation of O* (step 2). The potential needed for all the OER steps to be downhill is 1.71 V corresponding to a theoretical overpotential of 0.48 V.

Next, we study the OER on the 1/2 ML OH* and 1/2 ML O* covered surface. We consider the OER to happen on one of the tilted O* since the OER on the OH* requires a higher overpotential (Fig. \ref{fig:2O_2OH_taon_oer_OH}). We find an OOH* intermediate formed on this O* to be unstable and to dissociate into a O* and OH* bound to the same Ta. In other words, it is energetically more favorable to adsorb the OH* resulting from deprotonating the water molecule on the Ta site rather than to form the OOH*. We hence proceed with a different mechanism, in which the initial tilted O* remains passive and the OER proceeds by the conventional mechanism as if the Ta site was bare (Fig. \ref{fig:TaON_panel} c). The free energy diagram reveals that the ODS is the formation of the OOH* with a free energy difference of 1.76 eV. This free energy change corresponds to an overpotential of 0.53 V, comparable to the one obtained on the clean surface. A recombination mechanism of the second O* with the tilted O* is highly unfavorable in all situations (see supporting information Section \ref{sec:SI_TaON}).

Finally, we consider the OER on the energetically most stable fully O*-covered 3 O* tilt surface (Fig. \ref{fig:TaON_full_figures} b). As discussed above, the states associated with the tilted and upright O* are located at very different energies and we thus consider these sites independently for the OER. Just as for the 1/2 ML O* and 1/2 ML OH* covered surface, we find that the OOH* intermediate formed on a tilted O* is unstable and dissociates into a O* and OH* bound to the same Ta and therefore proceed with the mechanism described above. The free energy diagram (Fig. \ref{fig:TaON_panel} d) reveals that the limiting step in this case is the deprotonation of OH* (step 2). The potential required for the process to become thermodynamically favourable is 1.84 V, which represents an overpotential of 0.61 V. For the OER on the upright O* (Fig. \ref{fig:oer_taon_full_vert}) we find the OOH* intermediate to be stable and calculate the Gibbs free energy change for the conventional mechanism. The ODS is the formation of the OOH* intermediate with a free energy change of 2.61 eV corresponding to an overpotential of 1.38 V, which is much higher than the one obtained on the tilted O*. 

\subsection*{SrO-terminated (001) surface}

\begin{figure}
	\centering
	\includegraphics[width=0.9\columnwidth]{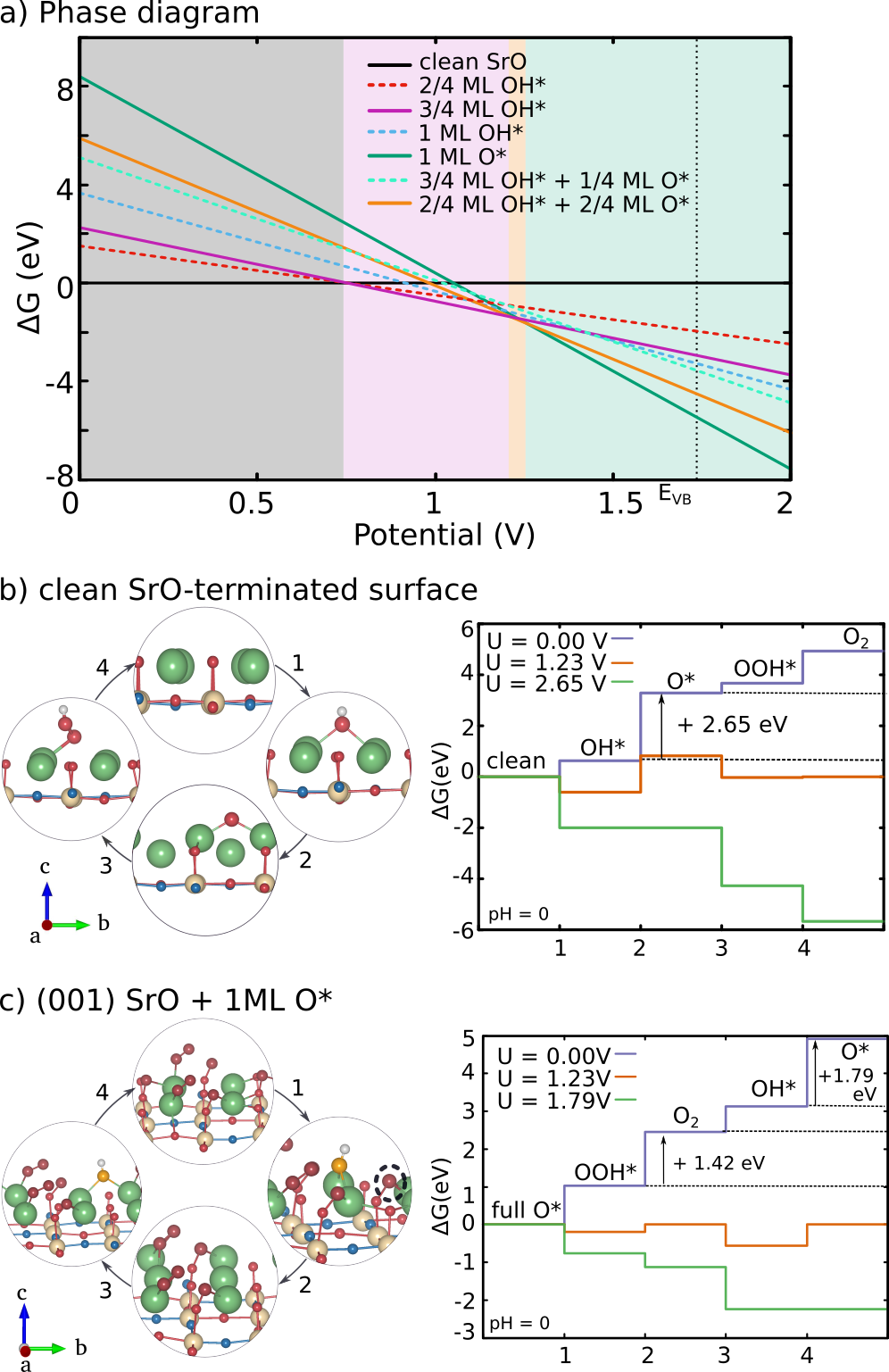}
	\caption{a) Surface Pourbaix diagram of the SrO-terminated surface as well as OER steps (left) and Gibbs free energy diagrams (right) for b) the clean and c) the fully O* covered surface.}
	\label{fig:SrO_panels}
\end{figure} 

Fig. \ref{fig:SrO_panels} a) shows the free energy relative to the clean surface for different adsorbate coverages of the SrO-terminated (001) surface as a function of the applied potential. For potentials lower than 0.75 V, the clean surface (with no adsorbates) is most stable, whereas for potentials in the range 0.75-1.2 V the surface covered with 3/4 ML OH* is the most stable, while the SrO surface covered with 1/2 ML of OH* and 1/2 ML of O* atoms becomes stable in a small range of 1.2-1.25 V. For even higher potentials the fully O*-covered surface is thermodynamically favoured.

We first investigate the OER on the clean surface (Fig. \ref{fig:SrO_panels} b). The OER proceeds by the conventional mechanism, with adsorbates in a bridge site between two Sr atoms, which is in agreement with other studies of O* adsorption on the A cations of perovskite oxide (001) surfaces \cite{akbay2016interaction, halwidl2016adsorption}. We note that the binding of the O* adsorbate results in a significant shift of the surface Sr atoms while the OOH* and OH* adsorbates do not lead to such a change in surface structure. This displacement of the surface atoms by the adsorbate is indicative of a low stability of the SrO-terminated surface in an oxidising environment and implies that this surface termination may not be very relevant for the OER. The free energy profile based on these potentially unstable structures (Fig. \ref{fig:SrO_panels} b) shows that the ODS corresponds to the formation of O* and has a free energy difference of 2.65 eV, corresponding to a theoretical overpotential of 1.42 V. This overpotential is significantly higher than on the TaON termination, which can be related to the too weak binding of the O* on the SrO termination. Indeed we find the O* binding energy on the SrO termination to be 0.73 eV more positive (weaker bond) than on the TaON surface (see Table \ref{tbl:table1} at the end of the article).

Next, we study the OER on the fully O*-covered surface (Fig. \ref{fig:SrO_panels} c) which is the most stable termination at application-relevant potentials > 1.23 V. In this configuration, two O* atoms recombine and form an O$_2$ molecule adsorbed on a Sr atom with the remaining two O* forming bonds with the surface O atoms. We note that this O$_2$ formation leads to a spurious energy lowering. We consider the OER on the still adsorbed O*. The mechanism that we find to be most favourable, does not involve OOH*, but proceeds by formation of OH* species in bridge positions between Sr sites and after deprotonation of OH* to O* their coupling with a pre-adsorbed O* (circled in Fig. \ref{fig:SrO_panels} c) and desorption of O$_2$, with a small energy difference of 1.42 eV. This mechanism has as the ODS the deprotonation of OH* with a free energy difference of 1.79 eV, corresponding to an overpotential of 0.67 V. The overpotential under these conditions is lower than on the clean SrO surface and comparable to the ones obtained for the TaON-terminated surface.  

\subsection*{(100) surface}

\begin{figure}
	\centering
	\includegraphics[width=0.9\columnwidth]{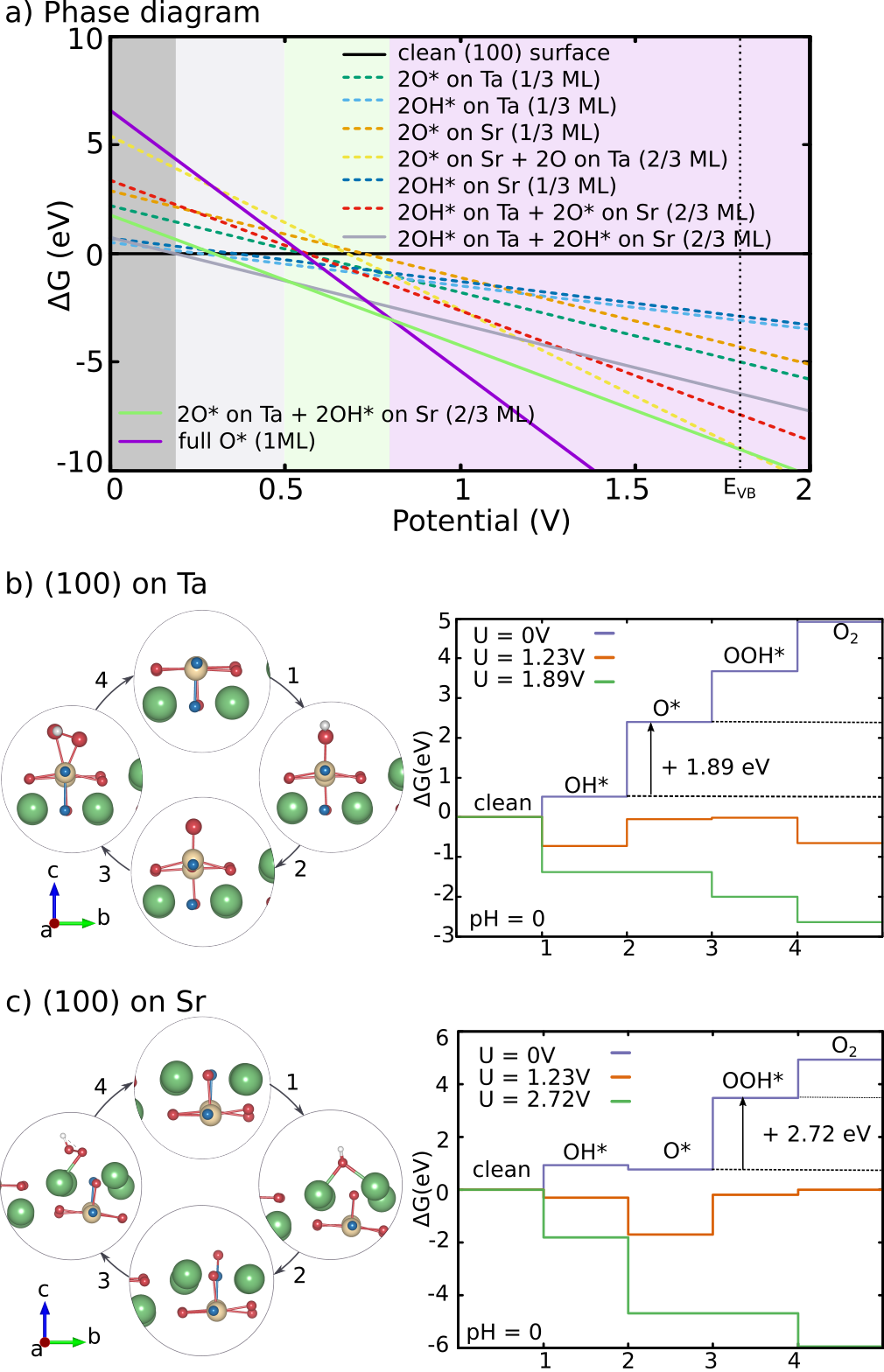}
	\caption{a) Surface Pourbaix diagram of the (100) surface (top) as well as OER steps (left) and the Gibbs free energy diagrams (right) for b) the Ta and c) the Sr site on the clean (100) surface.}
	\label{fig:100_panels}
\end{figure}

We further study the (100) surface of Sr$_2$TaO$_3$N, which exposes Sr and Ta reaction sites at a ratio of 4:2. We investigate different O* and OH* coverages on both sites (Fig. \ref{fig:100_panels} a) in order to determine the preferred coverage of the surface as a function of the potential. We see that the OH* covered termination (2/3 ML) becomes most stable for potentials between 0.18 V and 0.50 V, after which first the OH* adsorbed at the Ta site deprotonates, followed by the OH* at the Sr site for potentials beyond 0.8 V. On the resulting fully O*-covered surface, we however notice an unstable surface structure with large displacements of Sr atoms and desorption of O$_2$ and ON (supplementary Fig. \ref{fig:100_full_o}). We thus assume that this fully O*-covered (100) surface will reconstruct as discussed in the next section.

We investigate the OER on the clean (100) surface, treating the two reaction sites separately (Fig. \ref{fig:100_panels} b and c). Interestingly, we observe that despite the ODS being the same (step 2) on the Sr and Ta terminated clean (001) surfaces as determined above, on the (100) surface the ODS changes to step 3 for the Sr site. The overpotential on the Ta site (0.66 V) is significantly lower than on the Sr site (1.49 V), which agrees with the higher activity of the Ta site compared to the Sr site already found above for the (001) surface terminations. Still, the (100) surface requires higher overpotentials than the (001) surfaces, implying a lower OER reactivity.

\subsection*{Reconstructed grooved (100) surface}

\begin{figure}
	\centering
	\includegraphics[width=0.9\columnwidth]{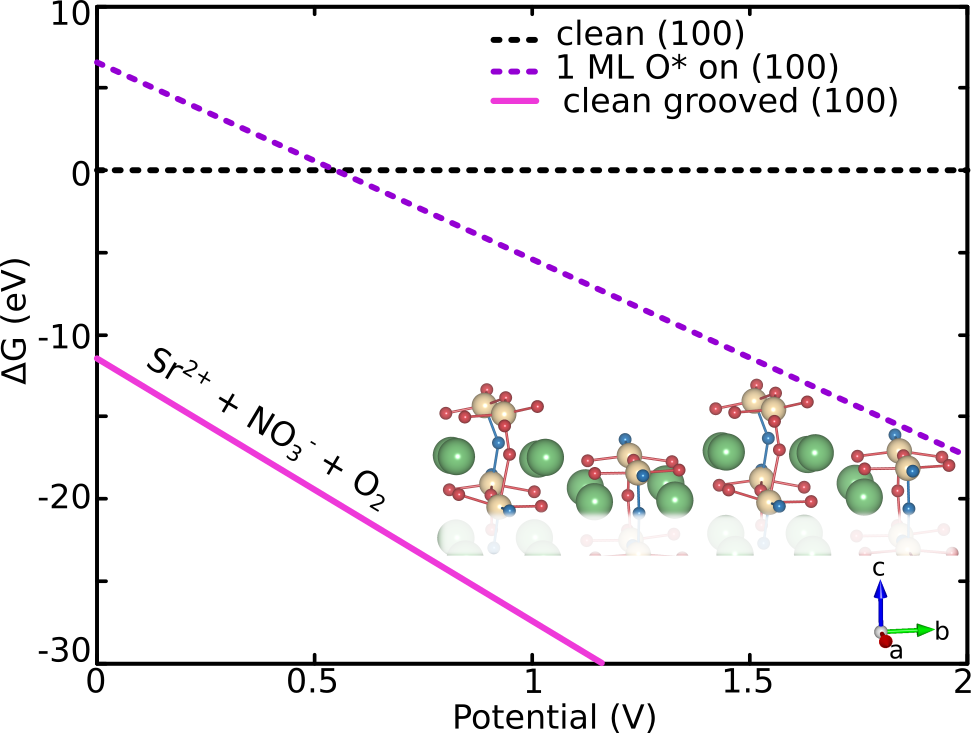}
	\caption{Stability of the O*-covered flat and grooved (100) surfaces with respect to the clean (100) surface. The structure of the grooved (100) surface is shown in the inset.}
	\label{fig:100_step_phase_diagram}
\end{figure}

As mentioned above, for potentials above 0.8 V, the fully O*-covered surface is the most stable but our calculations show an instability of the surface. We hence investigate the OER on a surface where the unstable Sr sites as well as their coordinating O and N atoms have been removed (see supporting information Section \ref{sec:SI_100}), resulting in a groove, while surface N atoms are substituted by O atoms to cancel polarity. The new surface shown in the inset of Fig. \ref{fig:100_step_phase_diagram}, exposes two Ta atoms on the upper terrace as reactive sites surrounded by O atoms and two Ta on the lower terrace surrounded by both N and O atoms.

\begin{figure}
	\centering
	\includegraphics[width=0.9\columnwidth]{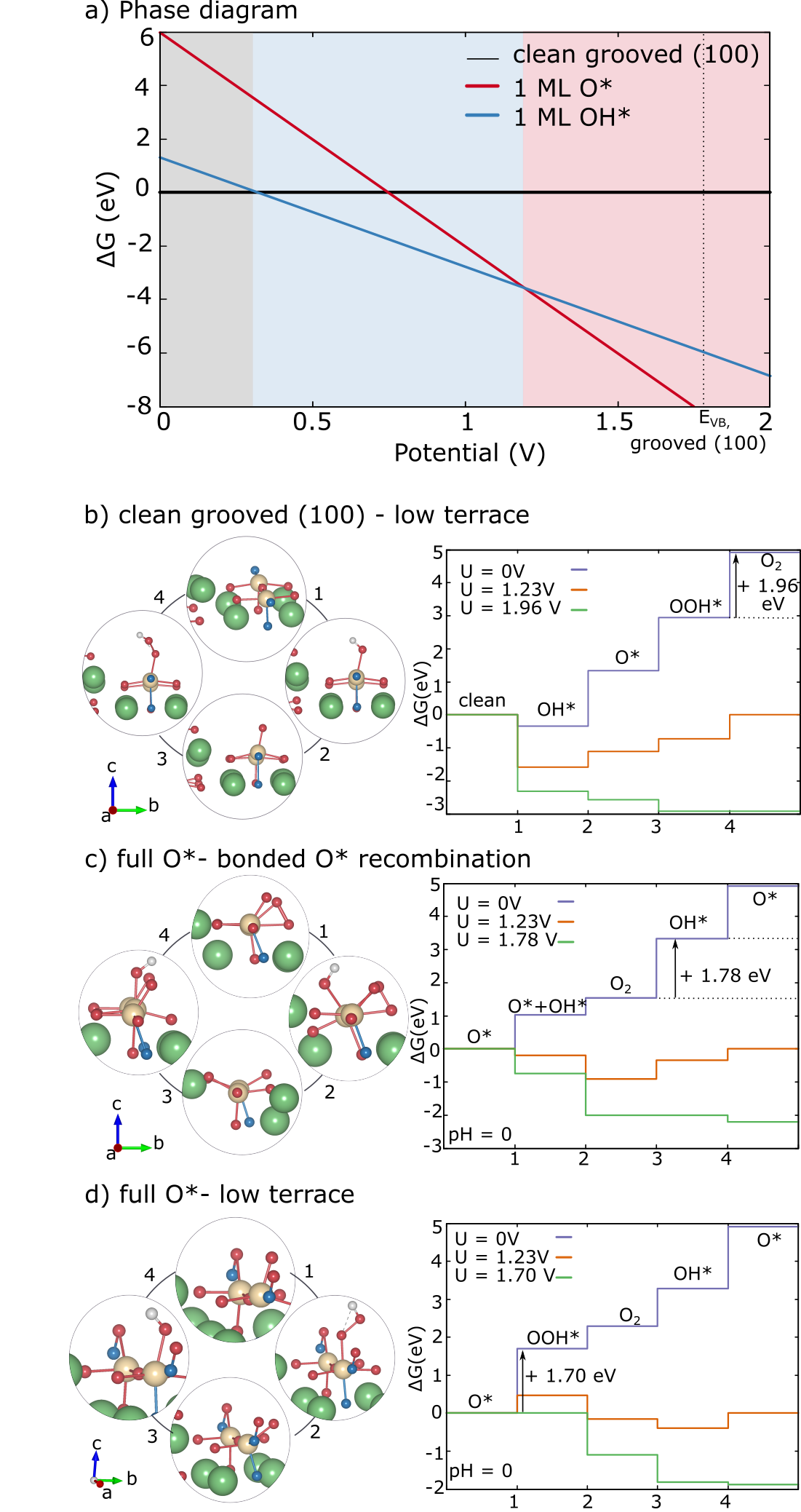}
	\caption{a) Surface Pourbaix diagram of the grooved (100) surface as well as OER steps (left) and Gibbs free energy diagram (right) for b) the clean and c)-d) different sites on the 1 ML-O*-covered surface.}
	\label{fig:100_step}
\end{figure}

We investigate the stability of the grooved surface with respect to the clean, non-reconstructed (100) surface (Fig. \ref{fig:100_step_phase_diagram}), considering that the removed Sr, N, and O atoms form solvated ions in water. Following the thermodynamic approach of Rong and Kolpak \cite{rong2015ab}, we calculate the free energy of the reconstructed (100) surface considering both the formation energy of the surface vacancies on the clean (100) surface and the formation energy of the solvated ions (see supporting information section \ref{sec:SI_methods}). Interestingly, we find that the reconstructed grooved surface is favoured over both the flat, clean and O*-covered surfaces at all potentials, showing the preference of the (100) surface to reconstruct in contact with water.

We first investigate the OER on the Ta-sites of both the upper and lower terrace of the grooved (100) surface without adsorbates (Fig. \ref{fig:100_step_overpotentials}) considering the conventional mechanism (results for other sites can be found in the supporting information). We find that the Ta-site of the lower terrace (bonded with a N atom along the \textit{z} direction) requires the lowest overpotential (Fig. \ref{fig:100_step} b) having as ODS the desorption of the O$_2$ molecule with a free energy difference of 1.96 eV corresponding to an overpotential of 0.73 V, higher than the one obtained on the clean TaON-terminated surface.

Next, we investigate the OER on the O*-covered termination, which we find to be more stable than the clean grooved surface for potentials higher than 1.19 V (Fig. \ref{fig:100_step} a). The two O* atoms on the upper terrace are adsorbed in a tilted configuration on the Ta atoms with one O* bonding with a surface O while the other does not form such a bond and we consider the OER on the two sites separately. We find the formation of OOH* to be unstable when O* is bonded with the surface O atom and hence consider the conventional mechanism on the over-coordinated Ta site similarly to the O*-covered TaON-terminated (001) surface (Fig. \ref{fig:100_step_full_O_mechanism}), for which we find an overpotential of 0.92 V. Interestingly, when considering an alternative mechanism involving the recombination of the O* and the deprotonated OH* intermediates and their desorption as an O$_2$ molecule, we find a lower overpotential (Fig. \ref{fig:100_step} c). The ODS for this mechanism is the adsorption of the second OH* intermediate on the empty Ta site with a free energy difference of 1.78 eV, corresponding to an overpotential of 0.55 V which is lower than the ones obtained on the (001) surfaces. We want to note that this recombination may however be associated with a kinetic barrier. On the O* that does not bind with surface O, we find the ODS to be the oxidation of OH* with a free energy change of 2.31 eV corresponding to an overpotential of 1.08 V (Fig. \ref{fig:100_step_vert_O_up}), which is larger than on the bonded O*, in agreement with the above results for the 1 ML O* covered TaON-terminated (001) surface.

We further investigate the OER on the lower terrace, which exposes one upright O* and one O* tilted and bonded with a surface N atom (Fig. \ref{fig:100_step} d). Interestingly, we find that the conventional mechanism on the upright O* results in a low overpotential of 0.47 V in contrast to the other upright cases. Here step 1 (OH* formation) is the ODS similarly to the case of the upright O* on the TaON-terminated surface. We can relate this to large structural relaxation (by 0.24 \AA) of the Ta atom along the \textit{c} direction, which results in the largest energy changes during step 1. We note that the recombination mechanism on the lower Ta-site has a significantly higher overpotential.

\subsection*{Positions of the valence-band maxima}

It is essential to determine the position of the valence-band edge relative to the normal hydrogen electrode (NHE) since this energy difference represents the OER bias potential resulting from the photo-excitation of electrons. We use an empirical method to estimate the band edges of Sr$_2$TaO$_3$N, which is based on the electronegativities of the contituent elements, $\chi_{element}$, and the bandgap, $E_{gap}$ of the bulk \cite{Castelli2013a}. The band edges for the RP oxynitride Sr$_2$TaO$_3$N within this approximation are:
\begin{equation}
E_{VB,CB} = E_0 +(\chi^2_{Sr}\chi_{Ta}\chi^3_{O}\chi_{N})^{1/7} \pm E_{gap}/2
\end{equation}
where $E_0$ is the difference between the vacuum level and the NHE (-4.5 eV). We previously determined $E_{gap}$ of bulk Sr$_2$TaO$_3$N as 2.005 eV using the hybrid functional HSE06 \cite{bouri2018bulk}. Using this value and the electronegativities calculated from the electron affinities and the ionization potentials as proposed by Mulliken \cite{mulliken1934new}, we find the VB and CB edges at 1.73 eV and -0.28 eV \textit{vs} NHE respectively. The VB edge that we are mostly interested in for water oxidation, is in agreement with the one previously determined based on G$_0$W$_0$ calculation of $E_{gap}$ and the same empirical method \cite{Castelli2013a}.

\begin{figure}
	\centering
	\includegraphics[width=0.9\columnwidth]{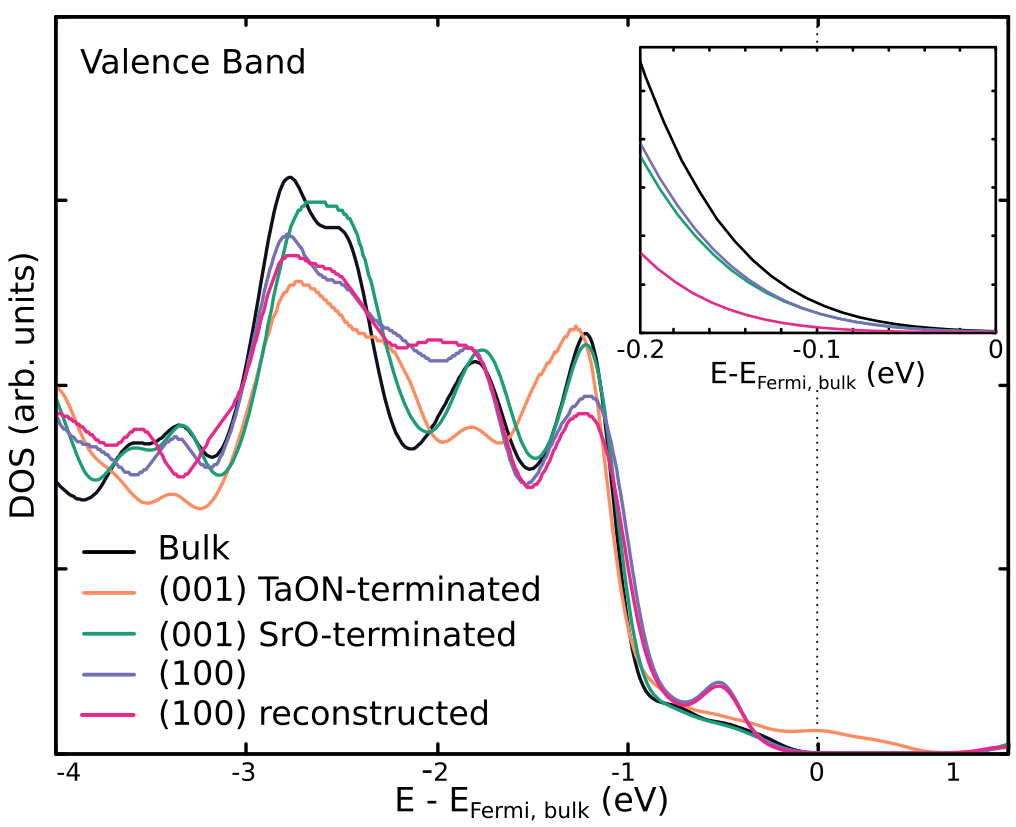}
	\caption{Electronic density of states (DOS) of bulk Sr$_2$TaO$_3$N and its surfaces, obtained using PBE. The DOS have been aligned at a Sr 4\textit{s} semi-core state -18.6 eV below the Fermi energy and each DOS is scaled according to the number of electrons in the respective structure.}
	\label{fig:dos_alignment}
\end{figure}

To further estimate the band edges of the various surfaces, we align the PBE density of states (DOS) of bulk Sr$_2$TaO$_3$N and its (001) and (100) surfaces at a Sr 4\textit{s} semi-core state (Fig. \ref{fig:dos_alignment}). We observe that the valence-band maximum (VBM) of both the SrO-terminated (001) surface (0.01 eV lower) and the (100) surface (0.025 eV lower) are in the same range as the one of the bulk while the VBM of the reconstructed (100) surface is 0.06 eV lower in energy. The VBM of the TaON-terminated (001) surface however is located at a significantly higher energy (0.665 eV) compared to all other VBMs due to the surface nitrogen states that are destabilised with respect to the bulk states \cite{bouri2018bulk}.

Since direct hybrid functional calculations of surface band edges are prohibitively expensive, we here consider the above values relative to the bulk hybrid functional VBM (1.73 eV) to estimate surface VBMs with respect to the NHE, which are 1.73 eV, 1.06 eV, 1.75 eV, and 1.78 eV for the SrO-terminated (001), the TaON-terminated (001), the (100) surface and the grooved (100) surface respectively. These VBM positions correspond to the potential vs. NHE that light irradiation provides and will determine the adsorbate coverage of the surface under illumination. We note that for all surfaces a full O* coverage is preferred at these potentials as indicated by the vertical $E_\mathrm{VB}$ line in the various surface Pourbaix diagrams. The difference between the respective VBM and the water oxidation potential (1.23 V vs. NHE) corresponds to the overpotential the surface provides. The TaON-terminated (001) surface does not provide a positive overpotential and is, despite it's promising catalytic activity, not able to drive the OER without an external potential. For the SrO-terminated (001) surface, the provided overpotential of 0.5 V is slightly smaller than the smallest computed OER overpotential of 0.56 V for the O*-covered surface, rendering this termination also unable to drive the OER. The (100) surface in its stable reconstructed form provides an overpotential of 0.55 V, which is larger than the smallest computed OER overpotential (0.47 V) and the reconstructed (100) surface can thus drive the OER.

\subsection*{Comparison of OER on (001) surfaces with layered and non-layered structure}

Experimental studies have compared the photocatalytic activity of layered with non-layered perovskite-oxides and have shown the higher photocatalytic activity of the layered materials \cite{Rodionov2016, Kudo2009, zeng2018phase}. In this section, we thus compare the overpotentials obtained for the (001) surfaces of layered Sr$_2$TaO$_3$N with the ones previously reported for non-layered SrTaO$_2$N (001) surfaces \cite{ouhbi2018water}. In that work a O*-covered surface with 4 O* tilted was considered but we find, in analogy to the above results, that a termination with 3 tilted and one upright O* is more favorable also for SrTaO$_2$N. In Table \ref{tbl:table1} we thus report, in addition to this 4 O* tilt case, the overpotential also for the energetically preferred 3 O* tilt structure.

\begin{table*}
\small
	\caption{Comparison of OER overpotentials, ODS and O binding energies of surfaces with layered and non-layered structure. For the 3 O* tilt structure, the results in parentheses correspond to the upright O* site.}
	\label{tbl:table1}
	\begin{tabular}{l|cc| cc| cc}
	\hline\hline
             & \multicolumn{2}{c|}{Overpotential} & \multicolumn{2}{c| }{ODS} & \multicolumn{2}{c}{O Binding Energy} \\
	\hline
	Surface termination & layered & non-layered & layered & non-layered & layered & non-layered\\
	\hline
	(001) SrO (clean)                 & 1.42       & 1.14\cite{ouhbi2018water} & 2    & 2\cite{ouhbi2018water} & 3.22 & 2.21\\
	(001) SrO (1 ML-O*)               & 0.56       & 1.14\cite{ouhbi2018water} & 4    & 2\cite{ouhbi2018water} & 2.24 & 2.03\\
	(001) TaON (clean)                & 0.48       & 1.01\cite{ouhbi2018water} & 2    & 3\cite{ouhbi2018water} & 2.50 & 0.62\\
	(001) TaON (1 ML-O*, 4 O* tilted) & 0.53       & 0.88\cite{ouhbi2018water} & 3    & 3\cite{ouhbi2018water} & 1.57 & 0.85\\
	(001) TaON (1 ML-O*, 3 O* tilted) & 0.61(1.38) & 0.52(1.26)                & 2(1) & 3(1)                   & 2.30 & 1.39\\	
	\hline\hline
\end{tabular}
\end{table*}

\begin{figure}
	\centering
	\includegraphics[width=0.9\columnwidth]{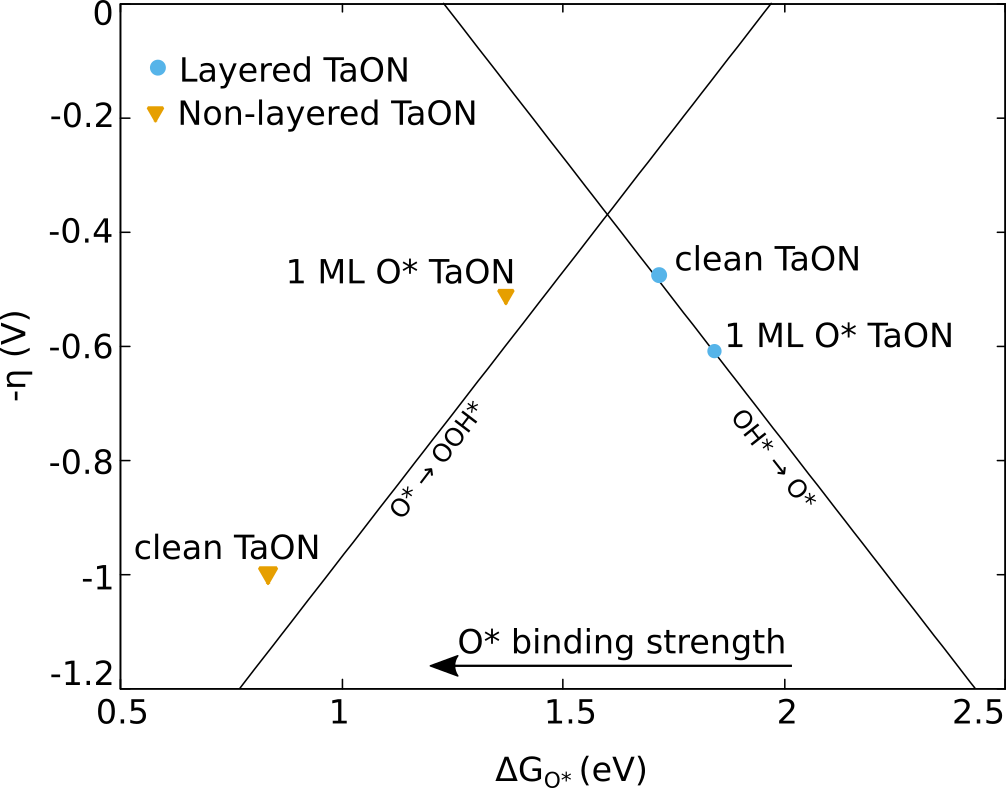}
	\caption{Volcano plot of the free energy difference of step 2 ($\mathrm{\Delta G_{O^*}}$) and the OER overpotential ($\eta$) for both layered and non-layered TaON-terminated (001) surfaces.}
	\label{fig:volcano}
\end{figure}

For the clean SrO-terminated (001) surface we predict a larger overpotential on the layered compared to the non-layered material even though the ODS is the same. Since the O* binding energy on the non-layered surface is 1.01 eV lower (stronger) than on the layered surface, the oxidation of OH* to O* requires less energy according to the universal scaling relations \cite{koper2013theory, man2011universality}. For the O*-covered SrO-terminated (001) surface, we find a different mechanism to be favored on the layered and the non-layered surface, which is reflected in the very different overpotentials, the layered material being significantly more active.

For the clean TaON-terminated (001) surface on the other hand we predict a significantly smaller OER overpotential for the layered surface compared to the non-layered one, the ODS being step 3 (OOH* formation) and step 2 (O* formation) respectively. This is due to the much stronger binding of O* to the non-layered surface, which, according to the commonly-used volcano analysis as a function of the free energy of step 2 \cite{man2011universality, rossmeisl2007electrolysis, Rossmeisl2005}, indicates that the two materials are located on different sides of the volcano tip. While the layered material is located on the right branch and limited by weak O* adsorption, the non-layered material is located on the left branch and limited by strong O* adsorption as shown in Fig. \ref{fig:volcano}. We find that a full ML O* adsorbate coverage increases the overpotential of the layered material, while it decreases the overpotential of the non-layered material. We can relate this to the fact that increasing the O* coverage decreases the binding strength of individual O* adsorbates. Given that the non-layered material is located on the left branch of the activity volcano, this change will shift the overpotential towards the tip of the volcano (Fig. \ref{fig:volcano}). The layered material, already being on the right branch, moves further away from the tip, resulting in a higher overpotential. For the fully O*-covered TaON-terminated (100) surface the trend of a higher activity of the layered material is retained only when considering the energetically less favorable 4 O* tilt surface, but it is inverted when considering the energetically favorable 3 O* tilt surface structure. This is due to the fact that in the 4 O* tilted structure O* is bound significantly stronger, which shifts this surface from the right to the left branch of the volcano.

\section*{Conclusions}

We investigated the surface adsorbate coverage and the OER mechanisms on the (001) and (100) surfaces of Sr$_2$TaO$_3$N. For all surfaces, we predict full coverage with O* adsorbates under light irradiation. We observe complex surface structures at high O* coverages with O* forming bonds with surface N atoms. 

The potential provided by the valence-band edge of the TaON-terminated (001) surface is more negative than the lowest OER potential and hence not sufficient to drive the reaction. The same is true for the SrO-terminated (001) surface, which is hence also unable to drive the OER.

For the (100) surface on the other hand, we predict a grooved reconstruction to be most stable that has a low overpotential of 0.47 V and provides a sufficient potential to drive the OER. Combined with the high carrier mobility perpendicular to the surface \cite{bouri2018bulk}, this renders the (100) surface the most relevant Sr$_2$TaO$_3$N surface to photocatalyze the OER.

Comparing the reconstructed grooved (100) surface of the layered Sr$_2$TaO$_3$N with the TaON-terminated (001) surface of non-layered SrTaO$_2$N, we find the smallest overpotential for the Ta site on the lower terrace of the former, which is in agreement with the improved catalytic activity of layered perovskite materials. 

\FloatBarrier
\bibliography{library}

\begin{thebibliography}{5}%
\makeatletter
\providecommand \@ifxundefined [1]{%
 \@ifx{#1\undefined}
}%
\providecommand \@ifnum [1]{%
 \ifnum #1\expandafter \@firstoftwo
 \else \expandafter \@secondoftwo
 \fi
}%
\providecommand \@ifx [1]{%
 \ifx #1\expandafter \@firstoftwo
 \else \expandafter \@secondoftwo
 \fi
}%
\providecommand \natexlab [1]{#1}%
\providecommand \enquote  [1]{``#1''}%
\providecommand \bibnamefont  [1]{#1}%
\providecommand \bibfnamefont [1]{#1}%
\providecommand \citenamefont [1]{#1}%
\providecommand \href@noop [0]{\@secondoftwo}%
\providecommand \href [0]{\begingroup \@sanitize@url \@href}%
\providecommand \@href[1]{\@@startlink{#1}\@@href}%
\providecommand \@@href[1]{\endgroup#1\@@endlink}%
\providecommand \@sanitize@url [0]{\catcode `\\12\catcode `\$12\catcode
  `\&12\catcode `\#12\catcode `\^12\catcode `\_12\catcode `\%12\relax}%
\providecommand \@@startlink[1]{}%
\providecommand \@@endlink[0]{}%
\providecommand \url  [0]{\begingroup\@sanitize@url \@url }%
\providecommand \@url [1]{\endgroup\@href {#1}{\urlprefix }}%
\providecommand \urlprefix  [0]{URL }%
\providecommand \Eprint [0]{\href }%
\providecommand \doibase [0]{http://dx.doi.org/}%
\providecommand \selectlanguage [0]{\@gobble}%
\providecommand \bibinfo  [0]{\@secondoftwo}%
\providecommand \bibfield  [0]{\@secondoftwo}%
\providecommand \translation [1]{[#1]}%
\providecommand \BibitemOpen [0]{}%
\providecommand \bibitemStop [0]{}%
\providecommand \bibitemNoStop [0]{.\EOS\space}%
\providecommand \EOS [0]{\spacefactor3000\relax}%
\providecommand \BibitemShut  [1]{\csname bibitem#1\endcsname}%
\let\auto@bib@innerbib\@empty
\bibitem [{\citenamefont {Vald{\'{e}}s}\ \emph {et~al.}(2008)\citenamefont
  {Vald{\'{e}}s}, \citenamefont {Qu}, \citenamefont {Kroes}, \citenamefont
  {Rossmeisl},\ and\ \citenamefont {N{\o}rskov}}]{Valdes2008SI}%
  \BibitemOpen
  \bibfield  {author} {\bibinfo {author} {\bibfnamefont {{\'{A}}.}~\bibnamefont
  {Vald{\'{e}}s}}, \bibinfo {author} {\bibfnamefont {Z.~W.}\ \bibnamefont
  {Qu}}, \bibinfo {author} {\bibfnamefont {G.~J.}\ \bibnamefont {Kroes}},
  \bibinfo {author} {\bibfnamefont {J.}~\bibnamefont {Rossmeisl}}, \ and\
  \bibinfo {author} {\bibfnamefont {J.~K.}\ \bibnamefont {N{\o}rskov}},\ }\href
  {\doibase 10.1021/jp711929d} {\bibfield  {journal} {\bibinfo  {journal}
  {Journal of Physical Chemistry C}\ }\textbf {\bibinfo {volume} {112}},\
  \bibinfo {pages} {9872} (\bibinfo {year} {2008})}\BibitemShut {NoStop}%
\bibitem [{\citenamefont {Rong}\ and\ \citenamefont
  {Kolpak}(2015)}]{rong2015abSI}%
  \BibitemOpen
  \bibfield  {author} {\bibinfo {author} {\bibfnamefont {X.}~\bibnamefont
  {Rong}}\ and\ \bibinfo {author} {\bibfnamefont {A.~M.}\ \bibnamefont
  {Kolpak}},\ }\href {\doibase 10.1021/acs.jpclett.5b00509} {\bibfield
  {journal} {\bibinfo  {journal} {The journal of physical chemistry letters}\
  }\textbf {\bibinfo {volume} {6}},\ \bibinfo {pages} {1785} (\bibinfo {year}
  {2015})}\BibitemShut {NoStop}%
\bibitem [{\citenamefont {Wagman}(1982)}]{wagman1982nbs}%
  \BibitemOpen
  \bibfield  {author} {\bibinfo {author} {\bibfnamefont {D.~D.}\ \bibnamefont
  {Wagman}},\ }\href@noop {} {\emph {\bibinfo {title} {{The NBS tables of
  chemical thermodynamic properties: Selected values for inorganic and C$_1$
  and C$_2$ organic substances in SI units}}}}\ (\bibinfo  {publisher}
  {National Bureau of Standards},\ \bibinfo {year} {1982})\BibitemShut
  {NoStop}%
\bibitem [{\citenamefont {Takeno}(2005)}]{takeno2005atlas}%
  \BibitemOpen
  \bibfield  {author} {\bibinfo {author} {\bibfnamefont {N.}~\bibnamefont
  {Takeno}},\ }\href@noop {} {\bibfield  {journal} {\bibinfo  {journal}
  {Geological survey of Japan Open File Report}\ }\textbf {\bibinfo {volume}
  {419}},\ \bibinfo {pages} {102} (\bibinfo {year} {2005})}\BibitemShut
  {NoStop}%
\bibitem [{\citenamefont {Persson}\ \emph {et~al.}(2012)\citenamefont
  {Persson}, \citenamefont {Waldwick}, \citenamefont {Lazic},\ and\
  \citenamefont {Ceder}}]{persson2012prediction}%
  \BibitemOpen
  \bibfield  {author} {\bibinfo {author} {\bibfnamefont {K.~A.}\ \bibnamefont
  {Persson}}, \bibinfo {author} {\bibfnamefont {B.}~\bibnamefont {Waldwick}},
  \bibinfo {author} {\bibfnamefont {P.}~\bibnamefont {Lazic}}, \ and\ \bibinfo
  {author} {\bibfnamefont {G.}~\bibnamefont {Ceder}},\ }\href {\doibase
  10.1103/PhysRevB.85.235438} {\bibfield  {journal} {\bibinfo  {journal} {Phys.
  Rev. B}\ }\textbf {\bibinfo {volume} {85}},\ \bibinfo {pages} {235438}
  (\bibinfo {year} {2012})}\BibitemShut {NoStop}%
\end{thebibliography}%


\begin{thebibliography}{36}%
\makeatletter
\providecommand \@ifxundefined [1]{%
 \@ifx{#1\undefined}
}%
\providecommand \@ifnum [1]{%
 \ifnum #1\expandafter \@firstoftwo
 \else \expandafter \@secondoftwo
 \fi
}%
\providecommand \@ifx [1]{%
 \ifx #1\expandafter \@firstoftwo
 \else \expandafter \@secondoftwo
 \fi
}%
\providecommand \natexlab [1]{#1}%
\providecommand \enquote  [1]{``#1''}%
\providecommand \bibnamefont  [1]{#1}%
\providecommand \bibfnamefont [1]{#1}%
\providecommand \citenamefont [1]{#1}%
\providecommand \href@noop [0]{\@secondoftwo}%
\providecommand \href [0]{\begingroup \@sanitize@url \@href}%
\providecommand \@href[1]{\@@startlink{#1}\@@href}%
\providecommand \@@href[1]{\endgroup#1\@@endlink}%
\providecommand \@sanitize@url [0]{\catcode `\\12\catcode `\$12\catcode
  `\&12\catcode `\#12\catcode `\^12\catcode `\_12\catcode `\%12\relax}%
\providecommand \@@startlink[1]{}%
\providecommand \@@endlink[0]{}%
\providecommand \url  [0]{\begingroup\@sanitize@url \@url }%
\providecommand \@url [1]{\endgroup\@href {#1}{\urlprefix }}%
\providecommand \urlprefix  [0]{URL }%
\providecommand \Eprint [0]{\href }%
\providecommand \doibase [0]{http://dx.doi.org/}%
\providecommand \selectlanguage [0]{\@gobble}%
\providecommand \bibinfo  [0]{\@secondoftwo}%
\providecommand \bibfield  [0]{\@secondoftwo}%
\providecommand \translation [1]{[#1]}%
\providecommand \BibitemOpen [0]{}%
\providecommand \bibitemStop [0]{}%
\providecommand \bibitemNoStop [0]{.\EOS\space}%
\providecommand \EOS [0]{\spacefactor3000\relax}%
\providecommand \BibitemShut  [1]{\csname bibitem#1\endcsname}%
\let\auto@bib@innerbib\@empty
\bibitem [{\citenamefont {Turner}(2004)}]{turner2004sustainable}%
  \BibitemOpen
  \bibfield  {author} {\bibinfo {author} {\bibfnamefont {J.~A.}\ \bibnamefont
  {Turner}},\ }\href {\doibase 10.1126/science.1103197} {\bibfield  {journal}
  {\bibinfo  {journal} {Science}\ }\textbf {\bibinfo {volume} {305}},\ \bibinfo
  {pages} {972} (\bibinfo {year} {2004})}\BibitemShut {NoStop}%
\bibitem [{\citenamefont {Gr{\"a}tzel}(2001)}]{Gratzel:79530}%
  \BibitemOpen
  \bibfield  {author} {\bibinfo {author} {\bibfnamefont {M.}~\bibnamefont
  {Gr{\"a}tzel}},\ }\href {\doibase 10.1038/35104607} {\bibfield  {journal}
  {\bibinfo  {journal} {Nature}\ }\textbf {\bibinfo {volume} {414}},\ \bibinfo
  {pages} {338} (\bibinfo {year} {2001})}\BibitemShut {NoStop}%
\bibitem [{\citenamefont {Rossmeisl}\ \emph {et~al.}(2005)\citenamefont
  {Rossmeisl}, \citenamefont {Logadottir},\ and\ \citenamefont
  {N{\o}rskov}}]{Rossmeisl2005}%
  \BibitemOpen
  \bibfield  {author} {\bibinfo {author} {\bibfnamefont {J.}~\bibnamefont
  {Rossmeisl}}, \bibinfo {author} {\bibfnamefont {A.}~\bibnamefont
  {Logadottir}}, \ and\ \bibinfo {author} {\bibfnamefont {J.~K.}\ \bibnamefont
  {N{\o}rskov}},\ }\href {\doibase 10.1016/j.chemphys.2005.05.038} {\bibfield
  {journal} {\bibinfo  {journal} {Chemical Physics}\ }\textbf {\bibinfo
  {volume} {319}},\ \bibinfo {pages} {178} (\bibinfo {year}
  {2005})}\BibitemShut {NoStop}%
\bibitem [{\citenamefont {Vald{\'{e}}s}\ \emph {et~al.}(2008)\citenamefont
  {Vald{\'{e}}s}, \citenamefont {Qu}, \citenamefont {Kroes}, \citenamefont
  {Rossmeisl},\ and\ \citenamefont {N{\o}rskov}}]{Valdes2008}%
  \BibitemOpen
  \bibfield  {author} {\bibinfo {author} {\bibfnamefont {{\'{A}}.}~\bibnamefont
  {Vald{\'{e}}s}}, \bibinfo {author} {\bibfnamefont {Z.~W.}\ \bibnamefont
  {Qu}}, \bibinfo {author} {\bibfnamefont {G.~J.}\ \bibnamefont {Kroes}},
  \bibinfo {author} {\bibfnamefont {J.}~\bibnamefont {Rossmeisl}}, \ and\
  \bibinfo {author} {\bibfnamefont {J.~K.}\ \bibnamefont {N{\o}rskov}},\ }\href
  {\doibase 10.1021/jp711929d} {\bibfield  {journal} {\bibinfo  {journal}
  {Journal of Physical Chemistry C}\ }\textbf {\bibinfo {volume} {112}},\
  \bibinfo {pages} {9872} (\bibinfo {year} {2008})}\BibitemShut {NoStop}%
\bibitem [{\citenamefont {Koper}(2013)}]{koper2013theory}%
  \BibitemOpen
  \bibfield  {author} {\bibinfo {author} {\bibfnamefont {M.~T.}\ \bibnamefont
  {Koper}},\ }\href {\doibase 10.1039/C3SC50205H} {\bibfield  {journal}
  {\bibinfo  {journal} {Chemical Science}\ }\textbf {\bibinfo {volume} {4}},\
  \bibinfo {pages} {2710} (\bibinfo {year} {2013})}\BibitemShut {NoStop}%
\bibitem [{\citenamefont {Man}\ \emph {et~al.}(2011)\citenamefont {Man},
  \citenamefont {Su}, \citenamefont {Calle-Vallejo}, \citenamefont {Hansen},
  \citenamefont {Mart{\'\i}nez}, \citenamefont {Inoglu}, \citenamefont
  {Kitchin}, \citenamefont {Jaramillo}, \citenamefont {N{\o}rskov},\ and\
  \citenamefont {Rossmeisl}}]{man2011universality}%
  \BibitemOpen
  \bibfield  {author} {\bibinfo {author} {\bibfnamefont {I.~C.}\ \bibnamefont
  {Man}}, \bibinfo {author} {\bibfnamefont {H.-Y.}\ \bibnamefont {Su}},
  \bibinfo {author} {\bibfnamefont {F.}~\bibnamefont {Calle-Vallejo}}, \bibinfo
  {author} {\bibfnamefont {H.~A.}\ \bibnamefont {Hansen}}, \bibinfo {author}
  {\bibfnamefont {J.~I.}\ \bibnamefont {Mart{\'\i}nez}}, \bibinfo {author}
  {\bibfnamefont {N.~G.}\ \bibnamefont {Inoglu}}, \bibinfo {author}
  {\bibfnamefont {J.}~\bibnamefont {Kitchin}}, \bibinfo {author} {\bibfnamefont
  {T.~F.}\ \bibnamefont {Jaramillo}}, \bibinfo {author} {\bibfnamefont {J.~K.}\
  \bibnamefont {N{\o}rskov}}, \ and\ \bibinfo {author} {\bibfnamefont
  {J.}~\bibnamefont {Rossmeisl}},\ }\href {\doibase 10.1002/cctc.201000397}
  {\bibfield  {journal} {\bibinfo  {journal} {ChemCatChem}\ }\textbf {\bibinfo
  {volume} {3}},\ \bibinfo {pages} {1159} (\bibinfo {year} {2011})}\BibitemShut
  {NoStop}%
\bibitem [{\citenamefont {Greeley}\ and\ \citenamefont
  {Markovic}(2012)}]{greeley2012road}%
  \BibitemOpen
  \bibfield  {author} {\bibinfo {author} {\bibfnamefont {J.}~\bibnamefont
  {Greeley}}\ and\ \bibinfo {author} {\bibfnamefont {N.~M.}\ \bibnamefont
  {Markovic}},\ }\href {\doibase 10.1039/C2EE21754F} {\bibfield  {journal}
  {\bibinfo  {journal} {Energy \& Environmental Science}\ }\textbf {\bibinfo
  {volume} {5}},\ \bibinfo {pages} {9246} (\bibinfo {year} {2012})}\BibitemShut
  {NoStop}%
\bibitem [{\citenamefont {Gim{\'e}nez}\ and\ \citenamefont
  {Bisquert}(2016)}]{gimenez2016photoelectrochemical}%
  \BibitemOpen
  \bibfield  {author} {\bibinfo {author} {\bibfnamefont {S.}~\bibnamefont
  {Gim{\'e}nez}}\ and\ \bibinfo {author} {\bibfnamefont {J.}~\bibnamefont
  {Bisquert}},\ }\href@noop {} {\  (\bibinfo {year} {2016})}\BibitemShut
  {NoStop}%
\bibitem [{\citenamefont {Kudo}\ and\ \citenamefont {Miseki}(2009)}]{Kudo2009}%
  \BibitemOpen
  \bibfield  {author} {\bibinfo {author} {\bibfnamefont {A.}~\bibnamefont
  {Kudo}}\ and\ \bibinfo {author} {\bibfnamefont {Y.}~\bibnamefont {Miseki}},\
  }\href {\doibase 10.1039/b800489g} {\bibfield  {journal} {\bibinfo  {journal}
  {Chemical Society reviews}\ }\textbf {\bibinfo {volume} {38}},\ \bibinfo
  {pages} {253} (\bibinfo {year} {2009})}\BibitemShut {NoStop}%
\bibitem [{\citenamefont {Takata}\ \emph {et~al.}(2015)\citenamefont {Takata},
  \citenamefont {Pan},\ and\ \citenamefont {Domen}}]{Takata2015}%
  \BibitemOpen
  \bibfield  {author} {\bibinfo {author} {\bibfnamefont {T.}~\bibnamefont
  {Takata}}, \bibinfo {author} {\bibfnamefont {C.}~\bibnamefont {Pan}}, \ and\
  \bibinfo {author} {\bibfnamefont {K.}~\bibnamefont {Domen}},\ }\href
  {\doibase 10.1088/1468-6996/16/3/033506} {\bibfield  {journal} {\bibinfo
  {journal} {Science and Technology of Advanced Materials}\ }\textbf {\bibinfo
  {volume} {16}},\ \bibinfo {pages} {033506} (\bibinfo {year}
  {2015})}\BibitemShut {NoStop}%
\bibitem [{\citenamefont {Fuertes}(2015)}]{Fuertes2015a}%
  \BibitemOpen
  \bibfield  {author} {\bibinfo {author} {\bibfnamefont {A.}~\bibnamefont
  {Fuertes}},\ }\href {\doibase 10.1039/C5MH00046G} {\bibfield  {journal}
  {\bibinfo  {journal} {Mater. Horiz.}\ }\textbf {\bibinfo {volume} {2}},\
  \bibinfo {pages} {453} (\bibinfo {year} {2015})}\BibitemShut {NoStop}%
\bibitem [{\citenamefont {Porter}\ \emph {et~al.}(2014)\citenamefont {Porter},
  \citenamefont {Huang},\ and\ \citenamefont {Woodward}}]{Porter2014}%
  \BibitemOpen
  \bibfield  {author} {\bibinfo {author} {\bibfnamefont {S.~H.}\ \bibnamefont
  {Porter}}, \bibinfo {author} {\bibfnamefont {Z.}~\bibnamefont {Huang}}, \
  and\ \bibinfo {author} {\bibfnamefont {P.~M.}\ \bibnamefont {Woodward}},\
  }\href {\doibase 10.1021/cg401230a} {\bibfield  {journal} {\bibinfo
  {journal} {Crystal Growth and Design}\ }\textbf {\bibinfo {volume} {14}},\
  \bibinfo {pages} {117} (\bibinfo {year} {2014})}\BibitemShut {NoStop}%
\bibitem [{\citenamefont {Ji}\ \emph {et~al.}(2005)\citenamefont {Ji},
  \citenamefont {Borse}, \citenamefont {Kim}, \citenamefont {Hwang},
  \citenamefont {Jang}, \citenamefont {Bae},\ and\ \citenamefont
  {Lee}}]{ji2005photocatalytic}%
  \BibitemOpen
  \bibfield  {author} {\bibinfo {author} {\bibfnamefont {S.~M.}\ \bibnamefont
  {Ji}}, \bibinfo {author} {\bibfnamefont {P.~H.}\ \bibnamefont {Borse}},
  \bibinfo {author} {\bibfnamefont {H.~G.}\ \bibnamefont {Kim}}, \bibinfo
  {author} {\bibfnamefont {D.~W.}\ \bibnamefont {Hwang}}, \bibinfo {author}
  {\bibfnamefont {J.~S.}\ \bibnamefont {Jang}}, \bibinfo {author}
  {\bibfnamefont {S.~W.}\ \bibnamefont {Bae}}, \ and\ \bibinfo {author}
  {\bibfnamefont {J.~S.}\ \bibnamefont {Lee}},\ }\href {\doibase
  10.1039/B417052K} {\bibfield  {journal} {\bibinfo  {journal} {Physical
  chemistry chemical physics}\ }\textbf {\bibinfo {volume} {7}},\ \bibinfo
  {pages} {1315} (\bibinfo {year} {2005})}\BibitemShut {NoStop}%
\bibitem [{\citenamefont {Ruddlesden}\ and\ \citenamefont
  {Popper}(1957)}]{ruddlesden1957}%
  \BibitemOpen
  \bibfield  {author} {\bibinfo {author} {\bibfnamefont {S.~N.}\ \bibnamefont
  {Ruddlesden}}\ and\ \bibinfo {author} {\bibfnamefont {P.}~\bibnamefont
  {Popper}},\ }\href {\doibase 10.1107/S0365110X57001929} {\bibfield  {journal}
  {\bibinfo  {journal} {Acta Crystallographica}\ }\textbf {\bibinfo {volume}
  {10}},\ \bibinfo {pages} {538} (\bibinfo {year} {1957})}\BibitemShut
  {NoStop}%
\bibitem [{\citenamefont {Dion}\ \emph {et~al.}(1981)\citenamefont {Dion},
  \citenamefont {Ganne},\ and\ \citenamefont {Tournoux}}]{dion1981nouvelles}%
  \BibitemOpen
  \bibfield  {author} {\bibinfo {author} {\bibfnamefont {M.}~\bibnamefont
  {Dion}}, \bibinfo {author} {\bibfnamefont {M.}~\bibnamefont {Ganne}}, \ and\
  \bibinfo {author} {\bibfnamefont {M.}~\bibnamefont {Tournoux}},\ }\href
  {\doibase 10.1016/0025-5408(81)90063-5} {\bibfield  {journal} {\bibinfo
  {journal} {Materials Research Bulletin}\ }\textbf {\bibinfo {volume} {16}},\
  \bibinfo {pages} {1429} (\bibinfo {year} {1981})}\BibitemShut {NoStop}%
\bibitem [{\citenamefont {Jacobson}\ \emph {et~al.}(1985)\citenamefont
  {Jacobson}, \citenamefont {Johnson},\ and\ \citenamefont
  {Lewandowski}}]{jacobson1985interlayer}%
  \BibitemOpen
  \bibfield  {author} {\bibinfo {author} {\bibfnamefont {A.}~\bibnamefont
  {Jacobson}}, \bibinfo {author} {\bibfnamefont {J.~W.}\ \bibnamefont
  {Johnson}}, \ and\ \bibinfo {author} {\bibfnamefont {J.}~\bibnamefont
  {Lewandowski}},\ }\href {\doibase 10.1021/ic00217a006} {\bibfield  {journal}
  {\bibinfo  {journal} {Inorganic Chemistry}\ }\textbf {\bibinfo {volume}
  {24}},\ \bibinfo {pages} {3727} (\bibinfo {year} {1985})}\BibitemShut
  {NoStop}%
\bibitem [{\citenamefont {Aurivillius}(1953)}]{aurivillius1953structure}%
  \BibitemOpen
  \bibfield  {author} {\bibinfo {author} {\bibfnamefont {B.}~\bibnamefont
  {Aurivillius}},\ }\href@noop {} {\bibfield  {journal} {\bibinfo  {journal}
  {Arkiv for Kemi}\ }\textbf {\bibinfo {volume} {5}},\ \bibinfo {pages} {39}
  (\bibinfo {year} {1953})}\BibitemShut {NoStop}%
\bibitem [{\citenamefont {Lichtenberg}\ \emph {et~al.}(2008)\citenamefont
  {Lichtenberg}, \citenamefont {Herrnberger},\ and\ \citenamefont
  {Wiedenmann}}]{lichtenberg2008synthesis}%
  \BibitemOpen
  \bibfield  {author} {\bibinfo {author} {\bibfnamefont {F.}~\bibnamefont
  {Lichtenberg}}, \bibinfo {author} {\bibfnamefont {A.}~\bibnamefont
  {Herrnberger}}, \ and\ \bibinfo {author} {\bibfnamefont {K.}~\bibnamefont
  {Wiedenmann}},\ }\href {\doibase 10.1016/j.progsolidstchem.2008.10.001}
  {\bibfield  {journal} {\bibinfo  {journal} {Progress in Solid State
  Chemistry}\ }\textbf {\bibinfo {volume} {36}},\ \bibinfo {pages} {253}
  (\bibinfo {year} {2008})}\BibitemShut {NoStop}%
\bibitem [{\citenamefont {Valdez}\ and\ \citenamefont
  {Spaldin}(2019)}]{valdez2019origin}%
  \BibitemOpen
  \bibfield  {author} {\bibinfo {author} {\bibfnamefont {M.~N.}\ \bibnamefont
  {Valdez}}\ and\ \bibinfo {author} {\bibfnamefont {N.~A.}\ \bibnamefont
  {Spaldin}},\ }\href {\doibase arXiv:1903.11847 [cond-mat.mtrl-sci]}
  {\bibfield  {journal} {\bibinfo  {journal} {arXiv:1903.11847
  [cond-mat.mtrl-sci]}\ } (\bibinfo {year} {2019}),\ arXiv:1903.11847
  [cond-mat.mtrl-sci]}\BibitemShut {NoStop}%
\bibitem [{\citenamefont {Castelli}\ \emph {et~al.}(2013)\citenamefont
  {Castelli}, \citenamefont {Garc{\'{i}}a-Lastra}, \citenamefont {H{\"{u}}ser},
  \citenamefont {Thygesen},\ and\ \citenamefont {Jacobsen}}]{Castelli2013a}%
  \BibitemOpen
  \bibfield  {author} {\bibinfo {author} {\bibfnamefont {I.~E.}\ \bibnamefont
  {Castelli}}, \bibinfo {author} {\bibfnamefont {J.~M.}\ \bibnamefont
  {Garc{\'{i}}a-Lastra}}, \bibinfo {author} {\bibfnamefont {F.}~\bibnamefont
  {H{\"{u}}ser}}, \bibinfo {author} {\bibfnamefont {K.~S.}\ \bibnamefont
  {Thygesen}}, \ and\ \bibinfo {author} {\bibfnamefont {K.~W.}\ \bibnamefont
  {Jacobsen}},\ }\href {\doibase 10.1088/1367-2630/15/10/105026} {\bibfield
  {journal} {\bibinfo  {journal} {New Journal of Physics}\ }\textbf {\bibinfo
  {volume} {15}},\ \bibinfo {pages} {105026} (\bibinfo {year}
  {2013})}\BibitemShut {NoStop}%
\bibitem [{\citenamefont {Bajdich}\ \emph {et~al.}(2013)\citenamefont
  {Bajdich}, \citenamefont {Garc{\'\i}a-Mota}, \citenamefont {Vojvodic},
  \citenamefont {N{\o}rskov},\ and\ \citenamefont
  {Bell}}]{bajdich2013theoretical}%
  \BibitemOpen
  \bibfield  {author} {\bibinfo {author} {\bibfnamefont {M.}~\bibnamefont
  {Bajdich}}, \bibinfo {author} {\bibfnamefont {M.}~\bibnamefont
  {Garc{\'\i}a-Mota}}, \bibinfo {author} {\bibfnamefont {A.}~\bibnamefont
  {Vojvodic}}, \bibinfo {author} {\bibfnamefont {J.~K.}\ \bibnamefont
  {N{\o}rskov}}, \ and\ \bibinfo {author} {\bibfnamefont {A.~T.}\ \bibnamefont
  {Bell}},\ }\href {\doibase 10.1021/ja405997s} {\bibfield  {journal} {\bibinfo
   {journal} {Journal of the American chemical Society}\ }\textbf {\bibinfo
  {volume} {135}},\ \bibinfo {pages} {13521} (\bibinfo {year}
  {2013})}\BibitemShut {NoStop}%
\bibitem [{\citenamefont {Montoya}\ \emph {et~al.}(2015)\citenamefont
  {Montoya}, \citenamefont {Garcia-Mota}, \citenamefont {N{\o}rskov},\ and\
  \citenamefont {Vojvodic}}]{montoya2015theoretical}%
  \BibitemOpen
  \bibfield  {author} {\bibinfo {author} {\bibfnamefont {J.~H.}\ \bibnamefont
  {Montoya}}, \bibinfo {author} {\bibfnamefont {M.}~\bibnamefont
  {Garcia-Mota}}, \bibinfo {author} {\bibfnamefont {J.~K.}\ \bibnamefont
  {N{\o}rskov}}, \ and\ \bibinfo {author} {\bibfnamefont {A.}~\bibnamefont
  {Vojvodic}},\ }\href {\doibase 10.1039/C4CP05259E} {\bibfield  {journal}
  {\bibinfo  {journal} {Physical Chemistry Chemical Physics}\ }\textbf
  {\bibinfo {volume} {17}},\ \bibinfo {pages} {2634} (\bibinfo {year}
  {2015})}\BibitemShut {NoStop}%
\bibitem [{\citenamefont {Ouhbi}\ and\ \citenamefont
  {Aschauer}(2018)}]{ouhbi2018water}%
  \BibitemOpen
  \bibfield  {author} {\bibinfo {author} {\bibfnamefont {H.}~\bibnamefont
  {Ouhbi}}\ and\ \bibinfo {author} {\bibfnamefont {U.}~\bibnamefont
  {Aschauer}},\ }\href {\doibase 10.1016/j.susc.2018.07.013} {\bibfield
  {journal} {\bibinfo  {journal} {Surface Science}\ }\textbf {\bibinfo {volume}
  {677}},\ \bibinfo {pages} {258 } (\bibinfo {year} {2018})}\BibitemShut
  {NoStop}%
\bibitem [{\citenamefont {Giannozzi}\ \emph {et~al.}(2009)\citenamefont
  {Giannozzi}, \citenamefont {Baroni}, \citenamefont {Bonini}, \citenamefont
  {Calandra}, \citenamefont {Car}, \citenamefont {Cavazzoni}, \citenamefont
  {Ceresoli}, \citenamefont {Chiarotti}, \citenamefont {Cococcioni},
  \citenamefont {Dabo}, \citenamefont {{Dal Corso}}, \citenamefont
  {de~Gironcoli}, \citenamefont {Fabris}, \citenamefont {Fratesi},
  \citenamefont {Gebauer}, \citenamefont {Gerstmann}, \citenamefont
  {Gougoussis}, \citenamefont {Kokalj}, \citenamefont {Lazzeri}, \citenamefont
  {Martin-Samos}, \citenamefont {Marzari}, \citenamefont {Mauri}, \citenamefont
  {Mazzarello}, \citenamefont {Paolini}, \citenamefont {Pasquarello},
  \citenamefont {Paulatto}, \citenamefont {Sbraccia}, \citenamefont {Scandolo},
  \citenamefont {Sclauzero}, \citenamefont {Seitsonen}, \citenamefont
  {Smogunov}, \citenamefont {Umari},\ and\ \citenamefont
  {Wentzcovitch}}]{Giannozzi2009}%
  \BibitemOpen
  \bibfield  {author} {\bibinfo {author} {\bibfnamefont {P.}~\bibnamefont
  {Giannozzi}}, \bibinfo {author} {\bibfnamefont {S.}~\bibnamefont {Baroni}},
  \bibinfo {author} {\bibfnamefont {N.}~\bibnamefont {Bonini}}, \bibinfo
  {author} {\bibfnamefont {M.}~\bibnamefont {Calandra}}, \bibinfo {author}
  {\bibfnamefont {R.}~\bibnamefont {Car}}, \bibinfo {author} {\bibfnamefont
  {C.}~\bibnamefont {Cavazzoni}}, \bibinfo {author} {\bibfnamefont
  {D.}~\bibnamefont {Ceresoli}}, \bibinfo {author} {\bibfnamefont {G.~L.}\
  \bibnamefont {Chiarotti}}, \bibinfo {author} {\bibfnamefont {M.}~\bibnamefont
  {Cococcioni}}, \bibinfo {author} {\bibfnamefont {I.}~\bibnamefont {Dabo}},
  \bibinfo {author} {\bibfnamefont {A.}~\bibnamefont {{Dal Corso}}}, \bibinfo
  {author} {\bibfnamefont {S.}~\bibnamefont {de~Gironcoli}}, \bibinfo {author}
  {\bibfnamefont {S.}~\bibnamefont {Fabris}}, \bibinfo {author} {\bibfnamefont
  {G.}~\bibnamefont {Fratesi}}, \bibinfo {author} {\bibfnamefont
  {R.}~\bibnamefont {Gebauer}}, \bibinfo {author} {\bibfnamefont
  {U.}~\bibnamefont {Gerstmann}}, \bibinfo {author} {\bibfnamefont
  {C.}~\bibnamefont {Gougoussis}}, \bibinfo {author} {\bibfnamefont
  {A.}~\bibnamefont {Kokalj}}, \bibinfo {author} {\bibfnamefont
  {M.}~\bibnamefont {Lazzeri}}, \bibinfo {author} {\bibfnamefont
  {L.}~\bibnamefont {Martin-Samos}}, \bibinfo {author} {\bibfnamefont
  {N.}~\bibnamefont {Marzari}}, \bibinfo {author} {\bibfnamefont
  {F.}~\bibnamefont {Mauri}}, \bibinfo {author} {\bibfnamefont
  {R.}~\bibnamefont {Mazzarello}}, \bibinfo {author} {\bibfnamefont
  {S.}~\bibnamefont {Paolini}}, \bibinfo {author} {\bibfnamefont
  {A.}~\bibnamefont {Pasquarello}}, \bibinfo {author} {\bibfnamefont
  {L.}~\bibnamefont {Paulatto}}, \bibinfo {author} {\bibfnamefont
  {C.}~\bibnamefont {Sbraccia}}, \bibinfo {author} {\bibfnamefont
  {S.}~\bibnamefont {Scandolo}}, \bibinfo {author} {\bibfnamefont
  {G.}~\bibnamefont {Sclauzero}}, \bibinfo {author} {\bibfnamefont {A.~P.}\
  \bibnamefont {Seitsonen}}, \bibinfo {author} {\bibfnamefont {A.}~\bibnamefont
  {Smogunov}}, \bibinfo {author} {\bibfnamefont {P.}~\bibnamefont {Umari}}, \
  and\ \bibinfo {author} {\bibfnamefont {R.~M.}\ \bibnamefont {Wentzcovitch}},\
  }\href {\doibase 10.1088/0953-8984/21/39/395502} {\bibfield  {journal}
  {\bibinfo  {journal} {Journal of physics. Condensed matter : an Institute of
  Physics journal}\ }\textbf {\bibinfo {volume} {21}},\ \bibinfo {pages}
  {395502} (\bibinfo {year} {2009})},\ \Eprint {http://arxiv.org/abs/0906.2569}
  {0906.2569} \BibitemShut {NoStop}%
\bibitem [{\citenamefont {Perdew}\ \emph {et~al.}(1996)\citenamefont {Perdew},
  \citenamefont {Burke},\ and\ \citenamefont {Ernzerhof}}]{Perdew1996}%
  \BibitemOpen
  \bibfield  {author} {\bibinfo {author} {\bibfnamefont {J.~P.}\ \bibnamefont
  {Perdew}}, \bibinfo {author} {\bibfnamefont {K.}~\bibnamefont {Burke}}, \
  and\ \bibinfo {author} {\bibfnamefont {M.}~\bibnamefont {Ernzerhof}},\ }\href
  {\doibase 10.1103/PhysRevLett.77.3865} {\bibfield  {journal} {\bibinfo
  {journal} {Physical Review Letters}\ }\textbf {\bibinfo {volume} {77}},\
  \bibinfo {pages} {3865} (\bibinfo {year} {1996})}\BibitemShut {NoStop}%
\bibitem [{\citenamefont {Vanderbilt}(1990)}]{Vanderbilt1990}%
  \BibitemOpen
  \bibfield  {author} {\bibinfo {author} {\bibfnamefont {D.}~\bibnamefont
  {Vanderbilt}},\ }\href {\doibase 10.1103/PhysRevB.41.7892} {\bibfield
  {journal} {\bibinfo  {journal} {Physical Review B}\ }\textbf {\bibinfo
  {volume} {41}},\ \bibinfo {pages} {7892} (\bibinfo {year}
  {1990})}\BibitemShut {NoStop}%
\bibitem [{\citenamefont {Monkhorst}\ and\ \citenamefont
  {Pack}(1976)}]{Pack1977}%
  \BibitemOpen
  \bibfield  {author} {\bibinfo {author} {\bibfnamefont {H.~J.}\ \bibnamefont
  {Monkhorst}}\ and\ \bibinfo {author} {\bibfnamefont {J.~D.}\ \bibnamefont
  {Pack}},\ }\href {\doibase 10.1103/PhysRevB.13.5188} {\bibfield  {journal}
  {\bibinfo  {journal} {Phys. Rev. B}\ }\textbf {\bibinfo {volume} {13}},\
  \bibinfo {pages} {5188} (\bibinfo {year} {1976})}\BibitemShut {NoStop}%
\bibitem [{\citenamefont {Bouri}\ and\ \citenamefont
  {Aschauer}(2018)}]{bouri2018bulk}%
  \BibitemOpen
  \bibfield  {author} {\bibinfo {author} {\bibfnamefont {M.}~\bibnamefont
  {Bouri}}\ and\ \bibinfo {author} {\bibfnamefont {U.}~\bibnamefont
  {Aschauer}},\ }\href {\doibase 10.1039/C7CP06791G} {\bibfield  {journal}
  {\bibinfo  {journal} {Physical Chemistry Chemical Physics}\ }\textbf
  {\bibinfo {volume} {20}},\ \bibinfo {pages} {2771} (\bibinfo {year}
  {2018})}\BibitemShut {NoStop}%
\bibitem [{\citenamefont {Bengtsson}(1999)}]{bengtsson1999dipole}%
  \BibitemOpen
  \bibfield  {author} {\bibinfo {author} {\bibfnamefont {L.}~\bibnamefont
  {Bengtsson}},\ }\href {\doibase 10.1103/PhysRevB.59.12301} {\bibfield
  {journal} {\bibinfo  {journal} {Physical Review B}\ }\textbf {\bibinfo
  {volume} {59}},\ \bibinfo {pages} {12301} (\bibinfo {year}
  {1999})}\BibitemShut {NoStop}%
\bibitem [{\citenamefont {Akbay}\ \emph {et~al.}(2016)\citenamefont {Akbay},
  \citenamefont {Staykov}, \citenamefont {Druce}, \citenamefont {T{\'e}llez},
  \citenamefont {Ishihara},\ and\ \citenamefont
  {Kilner}}]{akbay2016interaction}%
  \BibitemOpen
  \bibfield  {author} {\bibinfo {author} {\bibfnamefont {T.}~\bibnamefont
  {Akbay}}, \bibinfo {author} {\bibfnamefont {A.}~\bibnamefont {Staykov}},
  \bibinfo {author} {\bibfnamefont {J.}~\bibnamefont {Druce}}, \bibinfo
  {author} {\bibfnamefont {H.}~\bibnamefont {T{\'e}llez}}, \bibinfo {author}
  {\bibfnamefont {T.}~\bibnamefont {Ishihara}}, \ and\ \bibinfo {author}
  {\bibfnamefont {J.~A.}\ \bibnamefont {Kilner}},\ }\href {\doibase
  10.1039/C6TA02715F} {\bibfield  {journal} {\bibinfo  {journal} {Journal of
  Materials Chemistry A}\ }\textbf {\bibinfo {volume} {4}},\ \bibinfo {pages}
  {13113} (\bibinfo {year} {2016})}\BibitemShut {NoStop}%
\bibitem [{\citenamefont {Halwidl}\ \emph {et~al.}(2016)\citenamefont
  {Halwidl}, \citenamefont {St{\"o}ger}, \citenamefont {Mayr-Schm{\"o}lzer},
  \citenamefont {Pavelec}, \citenamefont {Fobes}, \citenamefont {Peng},
  \citenamefont {Mao}, \citenamefont {Parkinson}, \citenamefont {Schmid},
  \citenamefont {Mittendorfer}, \citenamefont {Redinger},\ and\ \citenamefont
  {Diebold}}]{halwidl2016adsorption}%
  \BibitemOpen
  \bibfield  {author} {\bibinfo {author} {\bibfnamefont {D.}~\bibnamefont
  {Halwidl}}, \bibinfo {author} {\bibfnamefont {B.}~\bibnamefont {St{\"o}ger}},
  \bibinfo {author} {\bibfnamefont {W.}~\bibnamefont {Mayr-Schm{\"o}lzer}},
  \bibinfo {author} {\bibfnamefont {J.}~\bibnamefont {Pavelec}}, \bibinfo
  {author} {\bibfnamefont {D.}~\bibnamefont {Fobes}}, \bibinfo {author}
  {\bibfnamefont {J.}~\bibnamefont {Peng}}, \bibinfo {author} {\bibfnamefont
  {Z.}~\bibnamefont {Mao}}, \bibinfo {author} {\bibfnamefont {G.~S.}\
  \bibnamefont {Parkinson}}, \bibinfo {author} {\bibfnamefont {M.}~\bibnamefont
  {Schmid}}, \bibinfo {author} {\bibfnamefont {F.}~\bibnamefont
  {Mittendorfer}}, \bibinfo {author} {\bibfnamefont {J.}~\bibnamefont
  {Redinger}}, \ and\ \bibinfo {author} {\bibfnamefont {U.}~\bibnamefont
  {Diebold}},\ }\href {\doibase 10.1038/nmat4512} {\bibfield  {journal}
  {\bibinfo  {journal} {Nature materials}\ }\textbf {\bibinfo {volume} {15}},\
  \bibinfo {pages} {450} (\bibinfo {year} {2016})}\BibitemShut {NoStop}%
\bibitem [{\citenamefont {Rong}\ and\ \citenamefont
  {Kolpak}(2015)}]{rong2015ab}%
  \BibitemOpen
  \bibfield  {author} {\bibinfo {author} {\bibfnamefont {X.}~\bibnamefont
  {Rong}}\ and\ \bibinfo {author} {\bibfnamefont {A.~M.}\ \bibnamefont
  {Kolpak}},\ }\href {\doibase 10.1021/acs.jpclett.5b00509} {\bibfield
  {journal} {\bibinfo  {journal} {The journal of physical chemistry letters}\
  }\textbf {\bibinfo {volume} {6}},\ \bibinfo {pages} {1785} (\bibinfo {year}
  {2015})}\BibitemShut {NoStop}%
\bibitem [{\citenamefont {Mulliken}(1934)}]{mulliken1934new}%
  \BibitemOpen
  \bibfield  {author} {\bibinfo {author} {\bibfnamefont {R.~S.}\ \bibnamefont
  {Mulliken}},\ }\href {\doibase 10.1063/1.1749394} {\bibfield  {journal}
  {\bibinfo  {journal} {The Journal of Chemical Physics}\ }\textbf {\bibinfo
  {volume} {2}},\ \bibinfo {pages} {782} (\bibinfo {year} {1934})}\BibitemShut
  {NoStop}%
\bibitem [{\citenamefont {Rodionov}\ and\ \citenamefont
  {Zvereva}(2016)}]{Rodionov2016}%
  \BibitemOpen
  \bibfield  {author} {\bibinfo {author} {\bibfnamefont {I.~A.}\ \bibnamefont
  {Rodionov}}\ and\ \bibinfo {author} {\bibfnamefont {I.~A.}\ \bibnamefont
  {Zvereva}},\ }\href {\doibase 10.1070/RCR4547} {\bibfield  {journal}
  {\bibinfo  {journal} {Russian Chemical Reviews}\ }\textbf {\bibinfo {volume}
  {85}},\ \bibinfo {pages} {248} (\bibinfo {year} {2016})}\BibitemShut
  {NoStop}%
\bibitem [{\citenamefont {Zeng}\ \emph {et~al.}(2018)\citenamefont {Zeng},
  \citenamefont {Bian}, \citenamefont {Cao}, \citenamefont {Ma}, \citenamefont
  {Liu}, \citenamefont {Zhu}, \citenamefont {Tan},\ and\ \citenamefont
  {Pan}}]{zeng2018phase}%
  \BibitemOpen
  \bibfield  {author} {\bibinfo {author} {\bibfnamefont {W.}~\bibnamefont
  {Zeng}}, \bibinfo {author} {\bibfnamefont {Y.}~\bibnamefont {Bian}}, \bibinfo
  {author} {\bibfnamefont {S.}~\bibnamefont {Cao}}, \bibinfo {author}
  {\bibfnamefont {Y.}~\bibnamefont {Ma}}, \bibinfo {author} {\bibfnamefont
  {Y.}~\bibnamefont {Liu}}, \bibinfo {author} {\bibfnamefont {A.}~\bibnamefont
  {Zhu}}, \bibinfo {author} {\bibfnamefont {P.}~\bibnamefont {Tan}}, \ and\
  \bibinfo {author} {\bibfnamefont {J.}~\bibnamefont {Pan}},\ }\href {\doibase
  10.1021/acsami.8b04837} {\bibfield  {journal} {\bibinfo  {journal} {ACS
  Applied Materials \& Interfaces}\ } (\bibinfo {year} {2018}),\
  10.1021/acsami.8b04837}\BibitemShut {NoStop}%
\bibitem [{\citenamefont {Rossmeisl}\ \emph {et~al.}(2007)\citenamefont
  {Rossmeisl}, \citenamefont {Qu}, \citenamefont {Zhu}, \citenamefont {Kroes},\
  and\ \citenamefont {N{\o}rskov}}]{rossmeisl2007electrolysis}%
  \BibitemOpen
  \bibfield  {author} {\bibinfo {author} {\bibfnamefont {J.}~\bibnamefont
  {Rossmeisl}}, \bibinfo {author} {\bibfnamefont {Z.-W.}\ \bibnamefont {Qu}},
  \bibinfo {author} {\bibfnamefont {H.}~\bibnamefont {Zhu}}, \bibinfo {author}
  {\bibfnamefont {G.-J.}\ \bibnamefont {Kroes}}, \ and\ \bibinfo {author}
  {\bibfnamefont {J.~K.}\ \bibnamefont {N{\o}rskov}},\ }\href {\doibase
  10.1016/j.jelechem.2006.11.008} {\bibfield  {journal} {\bibinfo  {journal}
  {Journal of Electroanalytical Chemistry}\ }\textbf {\bibinfo {volume}
  {607}},\ \bibinfo {pages} {83} (\bibinfo {year} {2007})}\BibitemShut
  {NoStop}%
\end{thebibliography}%

\clearpage

\setcounter{page}{1}
\renewcommand{\thetable}{S\arabic{table}}  
\setcounter{table}{0}
\renewcommand{\thefigure}{S\arabic{figure}}
\setcounter{figure}{0}
\renewcommand{\thesection}{S\arabic{section}}
\setcounter{section}{0}
\renewcommand{\theequation}{S\arabic{equation}}
\setcounter{equation}{0}
\onecolumngrid

\begin{center}
\textbf{Supplementary information for\\\vspace{0.5 cm}
\large Density functional theory study of the oxygen evolution activity on Sr$_2$TaO$_3$N surfaces\\\vspace{0.3 cm}}
Maria Bouri and Ulrich Aschauer

\small
\textit{Department of Chemistry and Biochemistry, University of Bern, Freiestrasse 3, CH-3012 Bern, Switzerland}\\\vspace{0.3 cm}
(Dated: \today)
\end{center}

\section{Methods}\label{sec:SI_methods}

We calculate the Gibbs free energy differences of the different surface terminations with respect to the clean (no adsorbates) surface:
\begin{equation}
\Delta{G} = \Delta{E}+(\Delta{ZPE}-T\Delta{S})-eU+k_{B} T ln10 \cdot pH \label{eq:dG}
\end{equation}
where $\Delta$E is the DFT total energy difference between the surface with adsorbates and without (clean), $\Delta$ZPE is the zero-point energy difference of the adsorbates and T$\Delta$S is the change in the entropy, both of which are calculated through phonon calculations (see Table \ref{tbl:tableZPE}), \text{$eU$} describes the energy shift due to the bias potential and the last term is the correction of the free energy of H$^+$ ions at pH$\neq$0.

\begin{table}[h]
	\caption{Values of the uero-point energy (ZPE) and change in entropy ($\Delta$S) determined by phonon calculations.}
	\label{tbl:tableZPE}
	\begin{tabular*}{0.25\textwidth}{c c c}
	\hline
	 & ZPE (eV) & TS (eV) \\
	\hline
	H$_2$O & 0.56 & 0.67 \\
	H$_2$ & 0.27 & 0.41 \\
	O* & 0.08 & 0.00 \\
	OH* & 0.33 & 0.00 \\
	OOH* & 0.43 & 0.00 \\
	O$_2$ & 0.10 & 0.64 \\	
\end{tabular*}
\end{table}

The OER mechanisms being studied in this work involve four concerted proton-coupled electron transfer steps (PCET) on the surfaces. The thermo-chemistry of this mechanism is investigated following the method developed by N{\o}rskov and coworkers \citeSI{Valdes2008SI}. The one-electron steps of the conventional OER mechanism (involving OOH*) are the following:
\begin{align}
\text{\hspace*{3cm}Step 1:\quad} &\mathrm{H_2O}(l) &+& \mathrm{^*} &\rightarrow\quad & \mathrm{HO^*} &+ \mathrm{H^+ + e^-}\hspace*{5cm}\\
\text{Step 2:\quad} &\mathrm{HO^*} &&&\rightarrow\quad & \mathrm{O^*} &+ \mathrm{H^+ + e^-} \hspace*{5cm}\\		
\text{Step 3:\quad} &\mathrm{H_2O}(l) &+& \mathrm{O^*} &\rightarrow\quad & \mathrm{HOO^*} &+ \mathrm{H^+ + e^-} \hspace*{5cm}\\ 
\text{Step 4:\quad} &\mathrm{HOO^*} &&&\rightarrow\quad & \mathrm{O_2} &+ \mathrm{H^+ + e^-} \hspace*{5cm}
\end{align}
where the symbol * represents a surface reaction site and O*, OH* and OOH* are adsorbed oxygen, hydroxyl and hydroperoxy groups respectively. We note that the above steps describe the OER mechanism under acidic conditions under which perovskite oxynitrides are unstable. These steps are however analogous to the ones under alkaline conditions by substituting $\text{H$_2$O(l)} \rightarrow \text{OH$^-$(aq) + H$^+$(aq)}$ and for convenience we work under acidic conditions. The Gibbs free energy difference for each of the above steps is calculated using equation \ref{eq:dG}, where $\Delta$E now is the DFT total energy difference between the two adsorbates. Due to the inability of DFT to accurately describe the O$_2$ molecule, its energy is calculated through the water dissociation reaction:
$\text{2H$_2$O } \rightarrow \text{ O$_2$ + 2H$_2$}$ considering the experimental free energy of dissociation of one water molecule (2.46 eV). In this case, we calculate: 
\begin{equation}
\Delta{G} = 4.92 eV = {E_{\mathrm{O}_2}}+2{E_{\mathrm{H}_2}}-2{E_{\mathrm{H_2O}}}+{(\Delta{ZPE}-T\Delta{S})}
\end{equation} 
Moreover, the free energy of the H$^+$ + e$^-$ is taken as equal to the energy of $\frac{1}{2}$H$_2$ in the gas phase at standard conditions (pH=0, \textit{p}=1 bar and T=298 K), which corresponds to the so-called computational standard hydrogen electrode (SHE). 

In the ideal case the minimum overpotential for this procedure would be accomplished by having the same Gibbs free energy difference (1.23 V) for each step. In real cases, an additional overpotential $\eta$ is needed for the OER to be thermodynamically favorable ($\Delta G < 0$):
\begin{equation}
\eta = max(\Delta{G}_1, \Delta{G}_2, \Delta{G}_3, \Delta{G}_4)/e -1.23 V
\end{equation}

Although the Gibbs free energy differences depend on the pH, the theoretical overpotential is independent of pH since both the free energy and the OER potential are affected by the same amount \text{k$_B$Tln10$\cdot$pH = 0.828 eV} at pH$\neq$0. For this reason, in this work we consider only pH=0.

The conventional OER mechanism on a fully O*-covered surface has been described by Vald{\'{e}}s \textit{et al.}\citeSI{Valdes2008SI}:
\begin{align}
\text{\hspace*{3cm}Step 1:\quad} &\mathrm{H_2O} &+& \mathrm{O^*} &\rightarrow\quad & \mathrm{HOO^*} &+ \mathrm{H^+ + e^-}\hspace*{5cm}\\
\text{Step 2:\quad} &\mathrm{HOO^*} &&&\rightarrow\quad & \mathrm{O_2} &+ \mathrm{H^+ + e^-} \hspace*{5cm}\\		
\text{Step 3:\quad} &\mathrm{H_2O} &+& \mathrm{^*} &\rightarrow\quad & \mathrm{HO^*} &+ \mathrm{H^+ + e^-} \hspace*{5cm}\\ 
\text{Step 4:\quad} &\mathrm{HO^*} &&&\rightarrow\quad & \mathrm{O^*} &+ \mathrm{H^+ + e^-} \hspace*{5cm}
\end{align}
where * indicates an adsorbed intermediate or an active site.

The binding energies of the intermediates OH*, O*, and OOH* are equal to:
\begin{align}
&\Delta{E}_{\mathrm{OH^*}} &&= E_{\mathrm{OH^*}} &- &E_* &- (E_{\mathrm{H_2O}} &- \frac{1}{2} E_{\mathrm{H}_2} )&\hspace*{2cm}\\
&\Delta{E}_{\mathrm{O^*}} &&= E_{\mathrm{O^*}} &- &E_* &- (E_{\mathrm{H_2O}} &- E_{\mathrm{H}_2} )&\hspace*{5cm}\\
&\Delta{E}_{\mathrm{OOH^*}} &&= E_{\mathrm{OOH^*}} &- &E_* &- (E_{\mathrm{H_2O}} &- \frac{3}{2} E_{\mathrm{H}_2} )&\hspace*{2cm}\\
\end{align}
where the energies $E_{\mathrm{OH*}}$, $E_{\mathrm{O*}}$, $E_{\mathrm{OOH*}}$ and $E_{\mathrm{*}}$ are DFT total energies of the surfaces with adsorbates OH*, O* and OOH* and the clean surface respectively.

For the formation of the reconstructed (100) surface, we consider the reaction of the (100) surface with water and the formation of solvated ions. We compare the stability of the grooved (100) surface with respect to the clean (100) surface following the method described in the work of Rong and Kolpak \citeSI{rong2015abSI} and calculate two energies $\Delta{G}_{1}$ and $\Delta{G}_{2}$. The first is the free energy required to remove one neutral atom A from the surface, which we calculate through DFT:
\begin{equation}
\Delta G_{1} = G_\mathrm{new} + \mu_{\mathrm{A}} - G_\mathrm{ref}
\end{equation}
where $G_\mathrm{new}$ is the free energy of the reconstructed surface and $G_\mathrm{ref}$ the one of the clean surface and $\mu$ is the chemical potential of atom A.

Next, we calculate the solvation free energy of atom A when it forms the solvated ion $H_{x}AO_{y}^{z-}$, which depends on the potential and pH:
\begin{equation}
\Delta G_{2} = \mu_{\mathrm{H_{x}AO_{y}^{z-}}} - \mu_{\mathrm{A}} - \sum n_{i}\mu_{i}
\end{equation}
where $\mu_{H_{x}AO_{y}^{z-}}$ is the chemical potential of the solvated ion and $n_{i}$ and $\mu_{i}$ are the number and the chemical potential of species $i=H_{2}O, H^{+}, e^{-}$. $\Delta G_{2}$ is calculated at the standard hydrogen electrode (SHE) and becomes:
\begin{equation}
\Delta G_{2} = \Delta G_\mathrm{SHE} - n_{e}(eU_\mathrm{SHE}) -2.3 n_{\mathrm{H}^{+}}kT\mathrm{pH} + kT\ln a_{\mathrm{H_{x}AO_{y}^{z-}}}
\end{equation}

The sum of these two energies results in the free energy of the reconstructed surface with respect to the clean surface. For the removal of more than one atom form the clean surface the free energy is: 
\begin{equation}
\Delta G =   (G_\mathrm{new} - G_\mathrm{ref} + \sum n_{\mathrm{A}}\mu_{\mathrm{A}}) + \sum n_{\mathrm{A}}[\Delta G_{2}]_{\mathrm{A}}
\end{equation}
where $n_\mathrm{A}$ is the number of atoms transferred to the solution. We note that all the free energies are calculated at pH=0 similar to the free energies of the O*/OH*-covered terminations.

In order to form the grooved (100) surface, we remove 4 Sr, 2 N and 1 O atoms from the clean (100) surface. $\Delta G$ is thus equal to:
\begin{equation}
\Delta G =   (G_\mathrm{stepped} - G_\mathrm{(100) clean} + 4  \mu_{\mathrm{Sr}} + 2 \mu_{\mathrm{N}} + \mu_{\mathrm{O}}) + 4 \Delta G_{2}^{\mathrm{Sr}^{2+}} + 2 \Delta G_{2}^{\mathrm{NH}^{4+},\mathrm{NO}^{3-},\mathrm{N}_{2}} + \Delta G_{2}^{\mathrm{O}}
\end{equation}
\qquad\qquad\qquad\qquad\qquad\qquad\qquad\qquad\qquad\qquad\qquad\qquad\qquad\qquad or
\begin{equation}
\Delta G =   (G_\mathrm{stepped} - G_\mathrm{(100) clean} + 4 E_{\mathrm{Sr}}^\mathrm{DFT} + 2 \frac{1}{2} E_{\mathrm{N}_2}^\mathrm{DFT} + \frac{1}{2} E_{\mathrm{O}_2}^\mathrm{DFT}) + 4 \Delta G_{2}^{\mathrm{Sr}^{2+}} + 2 \times \Delta G_{2}^{i} + \Delta G_{2}^{\mathrm{O}}
\end{equation}
where $i$ designates different solvated nitrogen species (see Table \ref{tbl:table3}). 

We report the DFT total energies of the different structures required to calculate $\Delta G_1$ and the formation energies of the formed solvated ions, which are taken from standard tables \citeSI{wagman1982nbs} and required to calculate $\Delta G_2$ in Tables \ref{tbl:table2} and \ref{tbl:table3} respectively. The Pourbaix diagram of nitrogen \citeSI{takeno2005atlas,persson2012prediction} shows that different ions are formed depending on the applied potential (at pH=0) and we thus consider each of them. The concentration for each atom we remove is equal to 1/16 ML.

\begin{table}[h]
\small
\centering
	\caption{DFT energies used to calculate $\Delta G_1$.}
	\label{tbl:table2}
	\begin{tabular}{l|r}
	\hline
	Structure & DFT energy (eV) \\
	\hline
	(100) clean surface & -143518.179 \\
	(100) stepped surface & -138535.569 \\
	O$_2$ & -867.870   \\
	N$_2$ & -541.334   \\	
	Sr bulk & -997.846   \\
	\hline
	\end{tabular}
\end{table}

\begin{table}[h]
\small
\centering
	\caption{Formation energies of solvated ion at pH=0.}
	\label{tbl:table3}
	\begin{tabular}{l|r |c}
	\hline
	 solvated ions & $\Delta G_f$\citeSI{wagman1982nbs,persson2012prediction} (eV) & $\Delta G_2$ (eV)\\
	\hline
	NH$_4^+$ (aq) & -0.793 & $+ eU + kT\ln a_{\mathrm{NH_4^+}} -0.793$\\
	NO$_3^-$ (aq) & -1.087 & $-5eU + kT\ln a_{\mathrm{NO_3^-}} -1.087$\\	
	NO$_2^-$ (aq) & -0.334 & $-3eU + kT\ln a_{\mathrm{NO_2^-}}  -0.334$\\
	N$_2$ (aq) & 0.188 & $0 eU - 0.188$\\
	N$_2$O$_2^{2-}$ (aq) & 1.438 & $-2eU + kT\ln a_{\mathrm{N_{2}O}_{2}^{2-}} + 1.438$\\
	HNO$_3$ (aq) & -1.146 & $-5eU -1.146$\\
	Sr$^{2+}$ (aq) & -5.595 & $-2eU + kT\ln a_{\mathrm{SrOH}^+} -5.595$\\
	O (aq) & -2.460 & $+2eU -2.46$ (O$_2$(g) as a reference)\\
	\hline
	\end{tabular}
\end{table}

\begin{figure}[h]
	\centering
	\includegraphics[width=0.5\columnwidth]{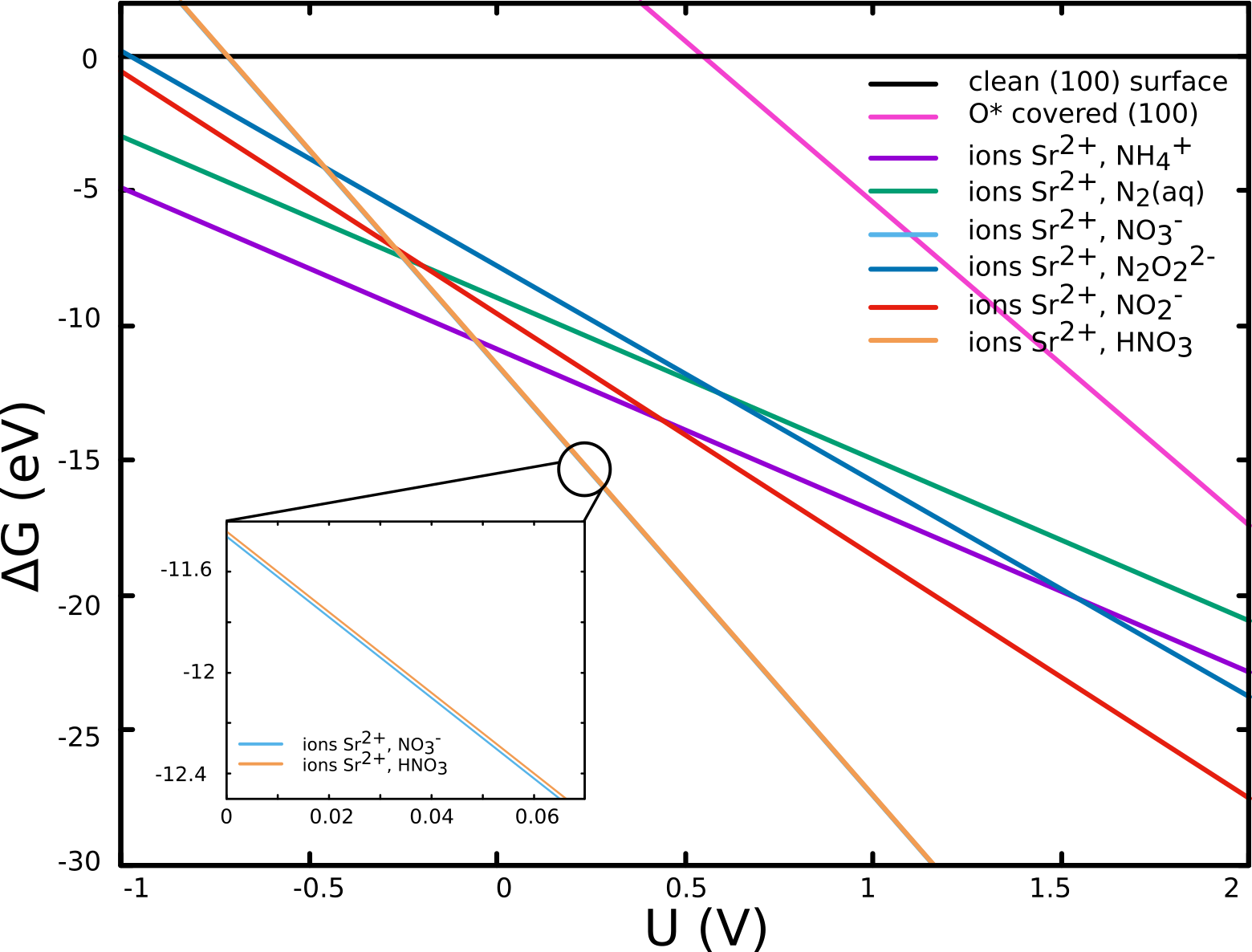}
	\caption{Phase stability of the stepped (100) surface with respect to the clean (100) surface considering the equilibrium with respect to different solvated nitrogen species.}
	\label{fig:step_100_ph_stab}
\end{figure}

The free energy of the stepped (100) surface with respect to the clean (100) surface considering the different nitrogen ions is shown in Fig. \ref{fig:step_100_ph_stab}. 

We determine the stability of the O*-covered and OH*-covered grooved (100) surfaces with respect to the clean grooved (100) surface at pH=0 in the same way we did for the (001) surfaces:
\begin{equation}
\Delta{G} = E_{4\mathrm{O}*} - E_{clean} + 4 E_{\mathrm{H}_{2}} - 4 E_{\mathrm{H_{2}O}} + (\Delta{ZPE}-T\Delta{S})-8 eU             
\end{equation}
\qquad\qquad\qquad\qquad\qquad\qquad\qquad\qquad\qquad\qquad\qquad\qquad\qquad and
\begin{equation}
\Delta{G} = E_{4\mathrm{OH}*} - E_{clean} + 2 E_{\mathrm{H}_{2}} - 4 E_{\mathrm{H_{2}O}} + (\Delta{ZPE}-T\Delta{S})-4 eU               .
\end{equation}

\FloatBarrier
\section{TaON-terminated (001) surface}\label{sec:SI_TaON}

\begin{figure}[h]
	\centering
	\includegraphics[width=0.4\columnwidth]{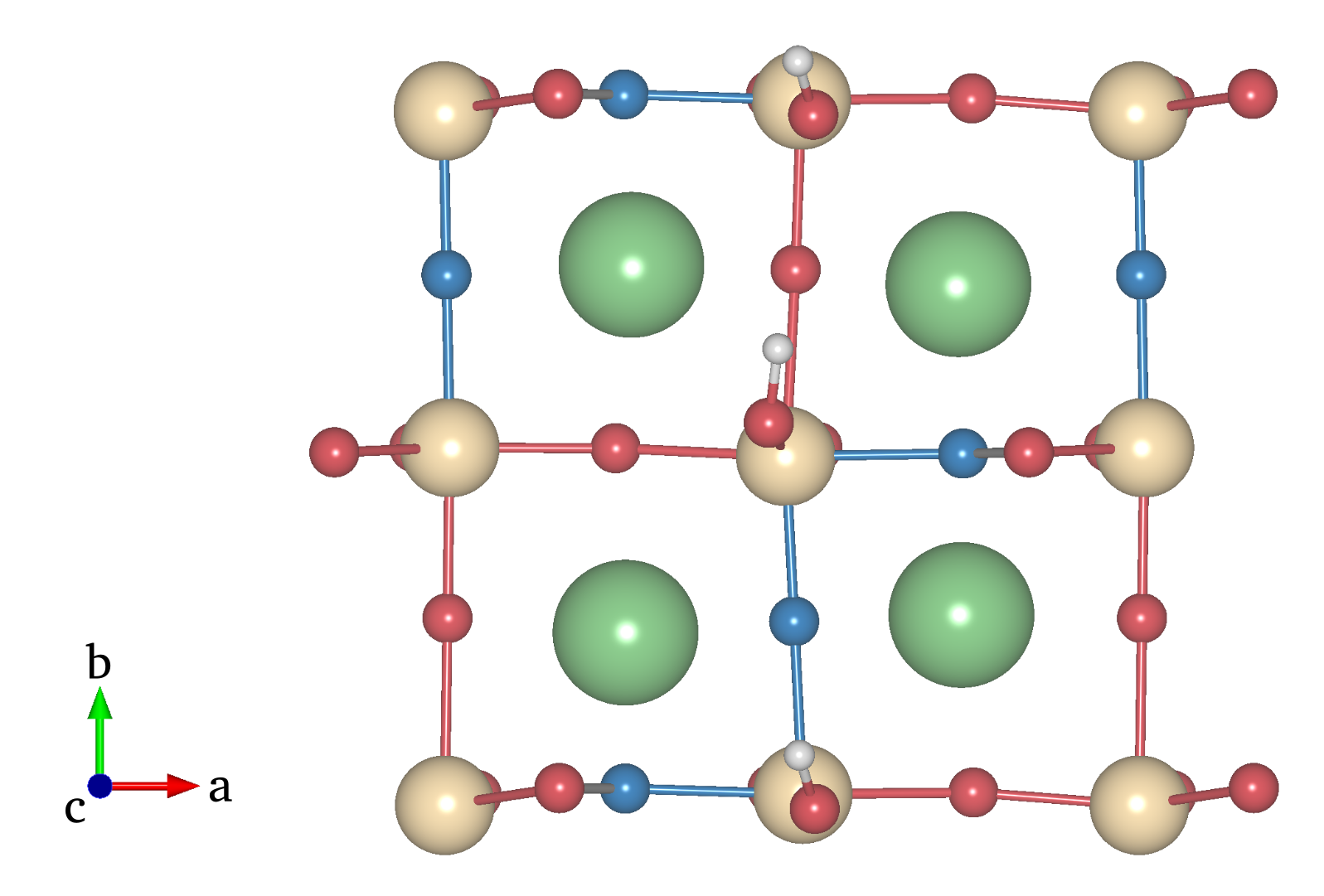}
	\caption{Configuration of the TaON-terminated (001) surface with 1/2 ML of O* and 1/2 ML of OH*.}
	\label{fig:2O_2OH_taon}
\end{figure}

The configuration with 1/2 ML of O* and 1/2 ML of OH* on the TaON-terminated (001) surface is shown in Fig. \ref{fig:2O_2OH_taon}. The different adsorbates are aligned parallel to the \textit{b} axis. The OH* are adsorbed vertically on top of the Ta-sites while the O* atoms are tilted on the surface and bonded also with surface N.  

\begin{figure}[h]
	\centering
	\includegraphics[width=0.53\columnwidth]{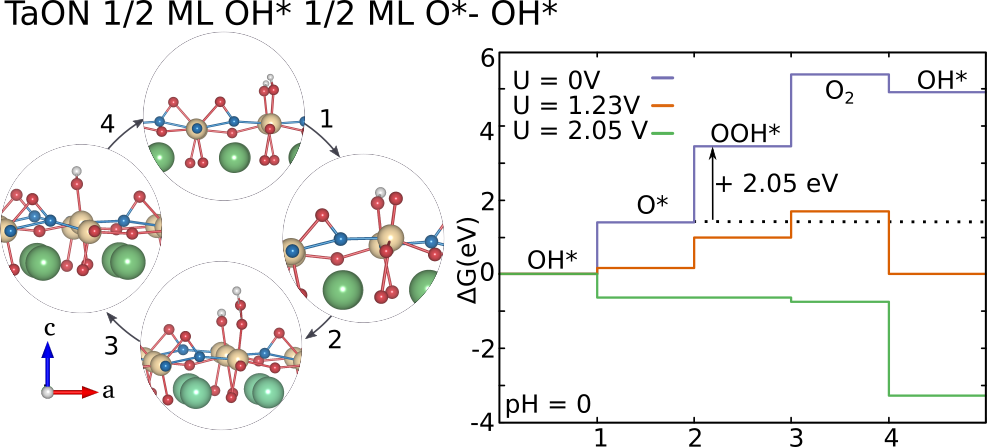}
	\caption{OER steps (left) and Gibbs free energy diagram (right) for the 1/2 ML OH* 1/ML O*-covered TaON-terminated (001) surface, considering OH* as reaction site.}
	\label{fig:2O_2OH_taon_oer_OH}
\end{figure}

\begin{figure}[h]
	\centering
	\includegraphics[width=0.8\columnwidth]{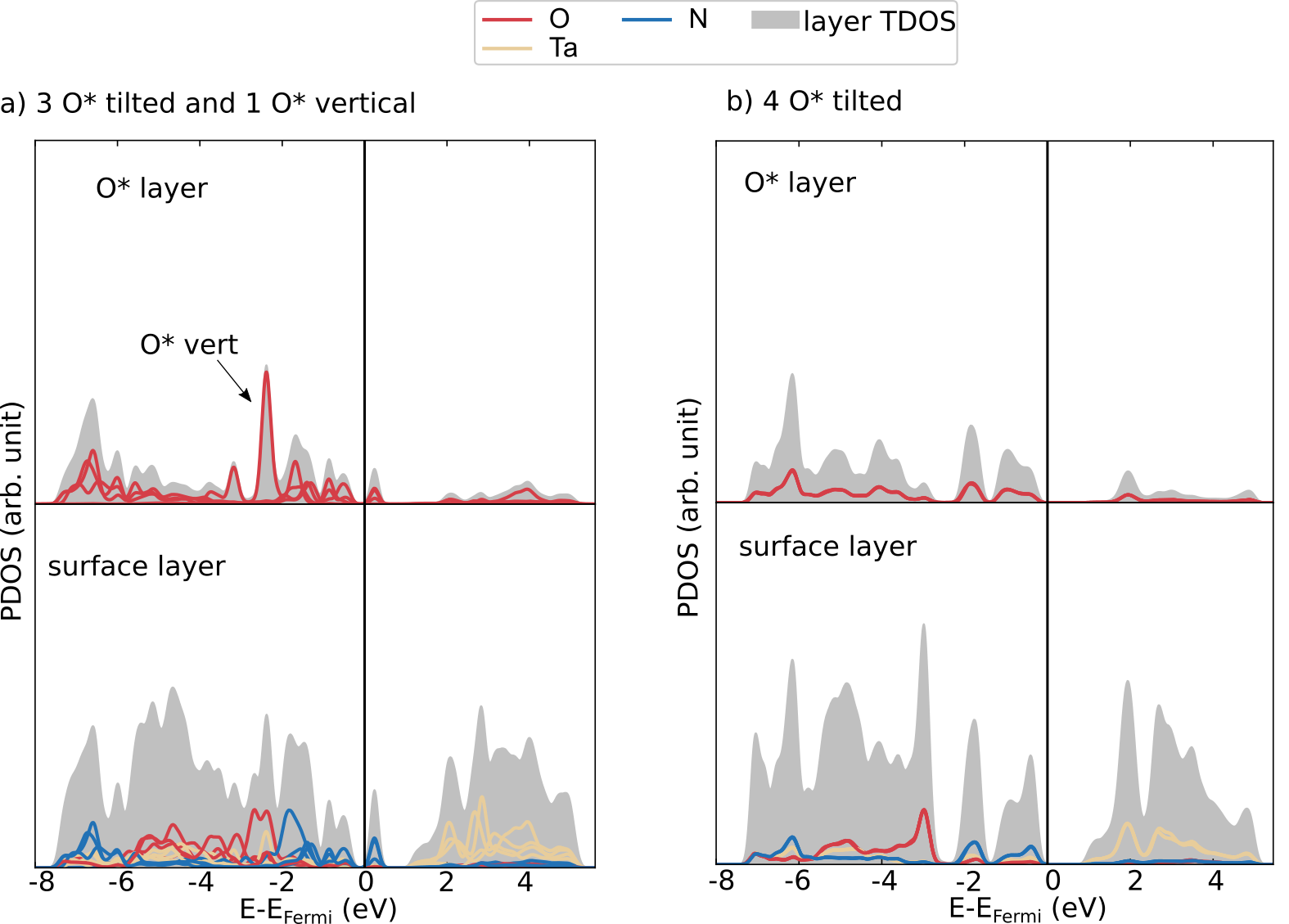}
	\caption{Density of states projected on the adsorbed O* atoms and the surface layer of (001) TaON-terminated surface with a) three O* tilted and one vertically adsorbed O* and b) all O* tilted on the surface.}
	\label{fig:pdos_O*_taon_surf_layer}
\end{figure}

\begin{figure}[h]
	\centering
	\includegraphics[width=0.6\columnwidth]{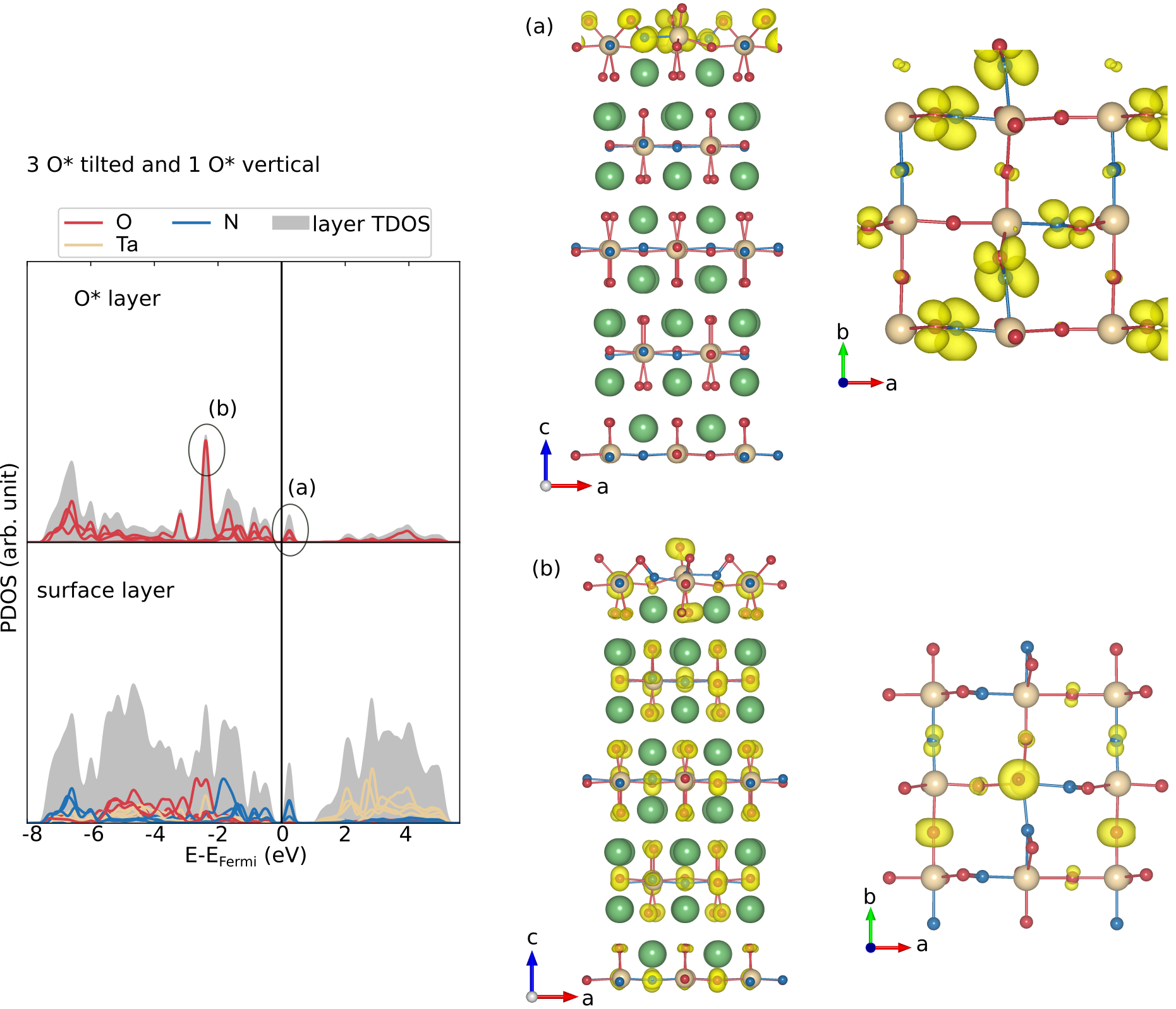}
	\caption{Integrated local density of states of a) the empty state formed in the band gap and b) the filled state in the VB with the highest contribution of oxygen states.}
	\label{fig:ildos_vb_peaks}
\end{figure}

In Fig. \ref{fig:pdos_O*_taon_surf_layer} we report the site-projected electronic density of states (DOS) of the O*-covered TaON-terminatd (001) surface in the 3 O* tilt and the 4 O* tilt configuration, while Fig. \ref{fig:ildos_vb_peaks} shows the integrated local density of states (ILDOS) of the indicated peaks of the 3 O* tilt configuration.

\begin{figure}[h]
	\centering
	\includegraphics[width=0.6\columnwidth]{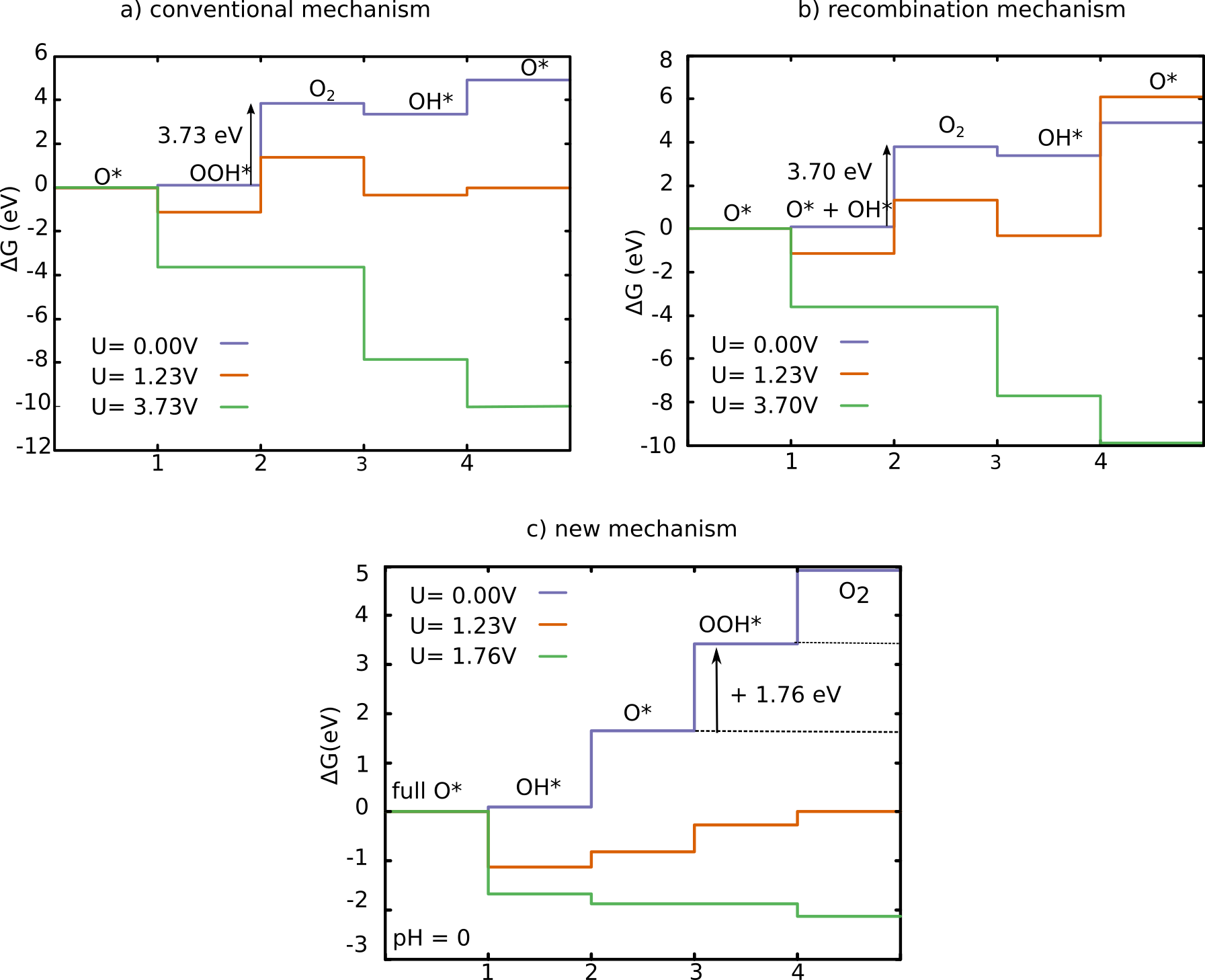}
	\caption{Free energy steps of a) the conventional mechanism and b) the new proposed mechanism on the 1 ML O* covered (001) TaON-terminated surface with all 4 O* tilted on the surface.}
	\label{fig:oer_full_tilt}
\end{figure}

We investigate the Gibbs free energy changes during the OER on the two energetically most favoured fully O*-covered TaON surfaces. We start with the less favorable case (4 O* tilt). The OER takes place on a O* atom. The conventional mechanism \citeSI{Valdes2008SI} (Fig. \ref{fig:oer_full_tilt} a) involves the adsorption of  OH* on the already adsorbed O* forming OOH*. During this conventional mechanism, we observe that the OOH* formation on the O*-covered TaON surface is particularly unstable and the OH fragment relaxes on the Ta site and not on the O* atom (OOH* dissociation). As a result the desorption of the OOH* and formation of the O$_2$ molecule has a high Gibbs free energy change (Fig. \ref{fig:oer_full_tilt} a) due to breaking of the strong Ta-O-N bond leading to a very high overpotential. We thus consider the recombination of the two O* to form O$_2$ that desorbs as follows:
\begin{align}
\text{\hspace*{3cm}Step 1:\quad} &\mathrm{H_2O} &+& \mathrm{O*} &\rightarrow\quad & \mathrm{HO^*} + \mathrm{O^*}  &+ \mathrm{H^+ + e^-}\hspace*{5cm}\\
\text{Step 2:\quad} &\mathrm{HO^*} &+& \mathrm{O*} &\rightarrow\quad & \mathrm{O_2} &+ \mathrm{H^+ + e^-} \hspace*{5cm}\\		
\text{Step 3:\quad} &\mathrm{H_2O} &+& \mathrm{*} &\rightarrow\quad & \mathrm{HO^*} &+ \mathrm{H^+ + e^-} \hspace*{5cm}\\ 
\text{Step 4:\quad} &\mathrm{HO^*} && &\rightarrow\quad & \mathrm{O^*} &+ \mathrm{H^+ + e^-} \hspace*{5cm}
\end{align}
We however find a high overpotential of 2.47 V for this mechanism with step 2 being the ODS (Fig. \ref{fig:oer_full_tilt} b). We conclude that neither the conventional nor the recombination mechanism are preferred on this surface. We also note that the recombination mechanism is unfavourable for other configurations of the O*-covered TaON-terminated surfaces as well as the SrO-terminated surface.   

Consequently, we suggest a different OER mechanism (Fig. \ref{fig:oer_full_tilt} c) in which the formation of the OOH* is stable in a further step. The OER steps are described below. The first water molecule enters and the OH fragment is adsorbed on the O* covered surface (step 1). The OOH* intermediate is not formed in this mechanism. It dissociates to O* and OH*, both adsorbed on the same Ta site. In step 2, the OH* is oxidized resulting in a Ta atom bonded with two O* atoms. The second water molecule deprotonates and the OH fragment is adsorbed on one of the two O* atoms (bonded with Ta) forming a stable OOH* intermediate (step 3). Then the OOH* deprotonates and desorbs as an O$_2$ molecule resulting in the initial O*-covered surface (step 4).  Overall, the overpotential is lower in this mechanism than in the conventional one rendering this process more favourable for this configuration of the O*-covered TaON surface.
\begin{align}
\text{\hspace*{3cm}Step 1:\quad} &\mathrm{H_2O} &+& \mathrm{O*} &\rightarrow\quad & \mathrm{HO^*} + \mathrm{O^*}  &+ \mathrm{H^+ + e^-}\hspace*{5cm}\\
\text{Step 2:\quad} &\mathrm{HO^*} &+& \mathrm{O*} &\rightarrow\quad & \mathrm{O^*} + \mathrm{O^*} &+ \mathrm{H^+ + e^-} \hspace*{5cm}\\		
\text{Step 3:\quad} &\mathrm{H_2O} &+& \mathrm{O^*} + \mathrm{O^*} &\rightarrow\quad & \mathrm{HOO^*} + \mathrm{O^*} &+ \mathrm{H^+ + e^-} \hspace*{5cm}\\ 
\text{Step 4:\quad} &\mathrm{HOO^*} &+& \mathrm{O^*} &\rightarrow\quad & \mathrm{O_2} + \mathrm{O^*} &+ \mathrm{H^+ + e^-} \hspace*{5cm}
\end{align}
From the Gibbs free-energy differences (Fig. \ref{fig:oer_full_tilt} c), we observe that the step with the highest free energy difference is the formation of the OOH*. The potential needed to make all the steps thermodynamically favourable is 1.76 V corresponding to a theoretical overpotential of 0.53 V, slightly higher than the one obtained in the clean TaON-terminated surface.

\begin{figure} [h]
	\centering
	\includegraphics[width=0.5\columnwidth]{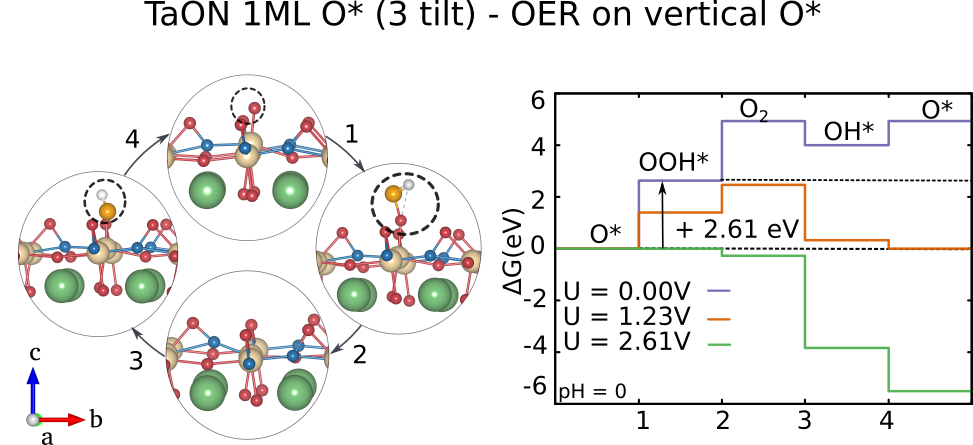}
	\caption{OER steps (left) and Gibbs free energy diagram (right) for the 1ML O*-covered 3 O* tilt TaON-terminated (001) surface, considering the upright O* as reaction site.}
	\label{fig:oer_taon_full_vert}
\end{figure}

On the 3 O* tilt surface the upright O* can also act as the active site. It is however associated with a high overpotential as shown in Fig. \ref{fig:oer_taon_full_vert}.

\FloatBarrier
\section{(001) SrO surface}

\begin{figure}[h]
	\centering
	\includegraphics[width=0.2\columnwidth]{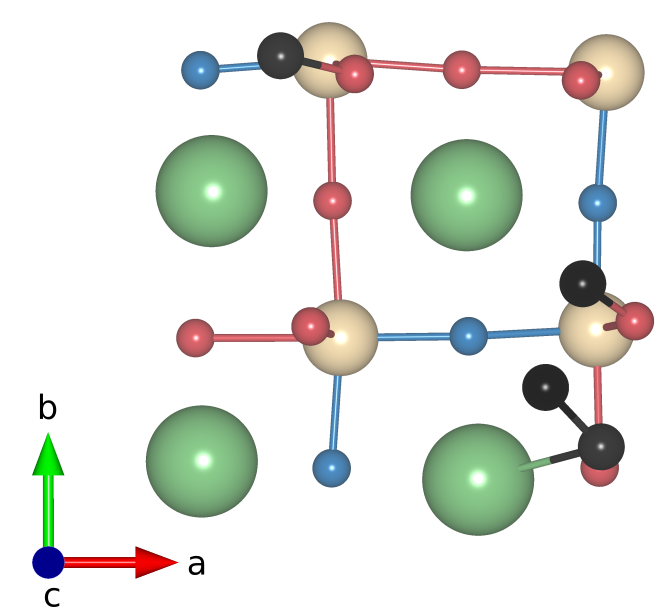}
	\caption{Representation of the 1 ML O* covered SrO-terminated (001) surface. The adsorbed O* atoms are drawn as black spheres.}
	\label{fig:001_full_o}
\end{figure}

In Fig. \ref{fig:001_full_o} we show a top view of the fully O*-covered SrO-terminated (001) surface.

\FloatBarrier
\section{Clean and grooved (100) surface}\label{sec:SI_100}

\begin{figure}[h]
	\centering
	\includegraphics[width=0.3\columnwidth]{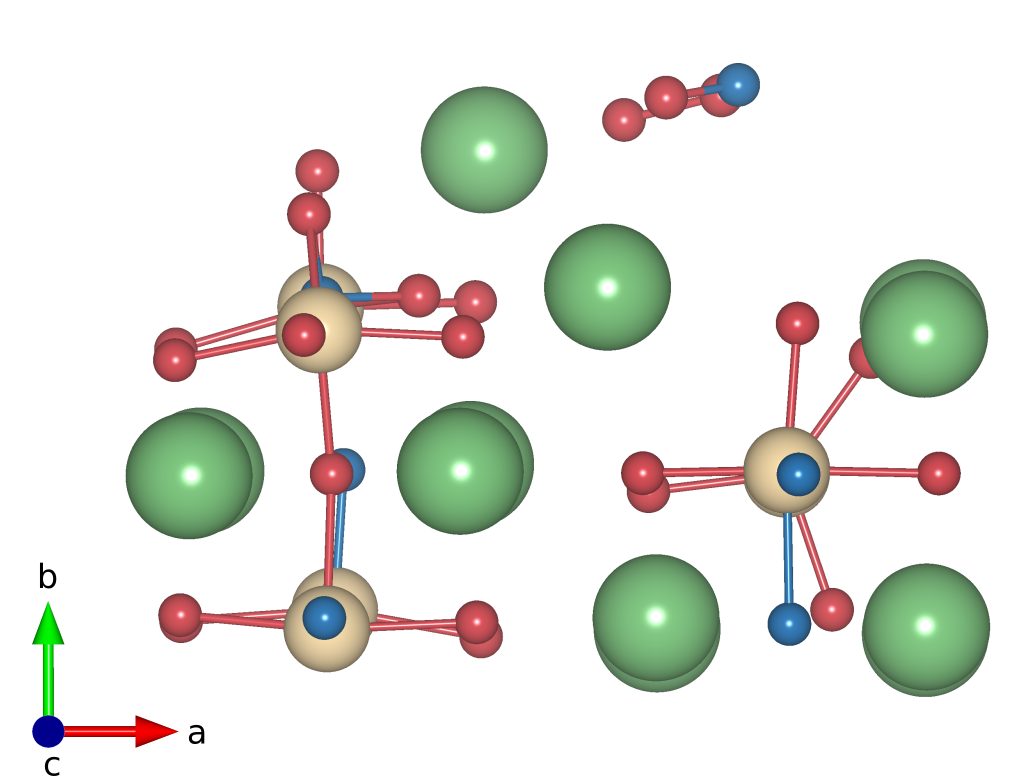}
	\caption{Structure of the 1 ML O* covered flat (100) surface.}
	\label{fig:100_full_o}
\end{figure}

For the (100) surface a full O* coverage on the Sr atoms leads to an unstable surface with adsorbed O* atoms shifting the surface Sr atoms and resulting in their desorption (Fig. \ref{fig:100_full_o}). We see that during relaxation the adsorbed O* atoms desorb from the surface together with surface N and Sr atoms.

\begin{figure}[h]
	\centering
	\includegraphics[width=0.5\columnwidth]{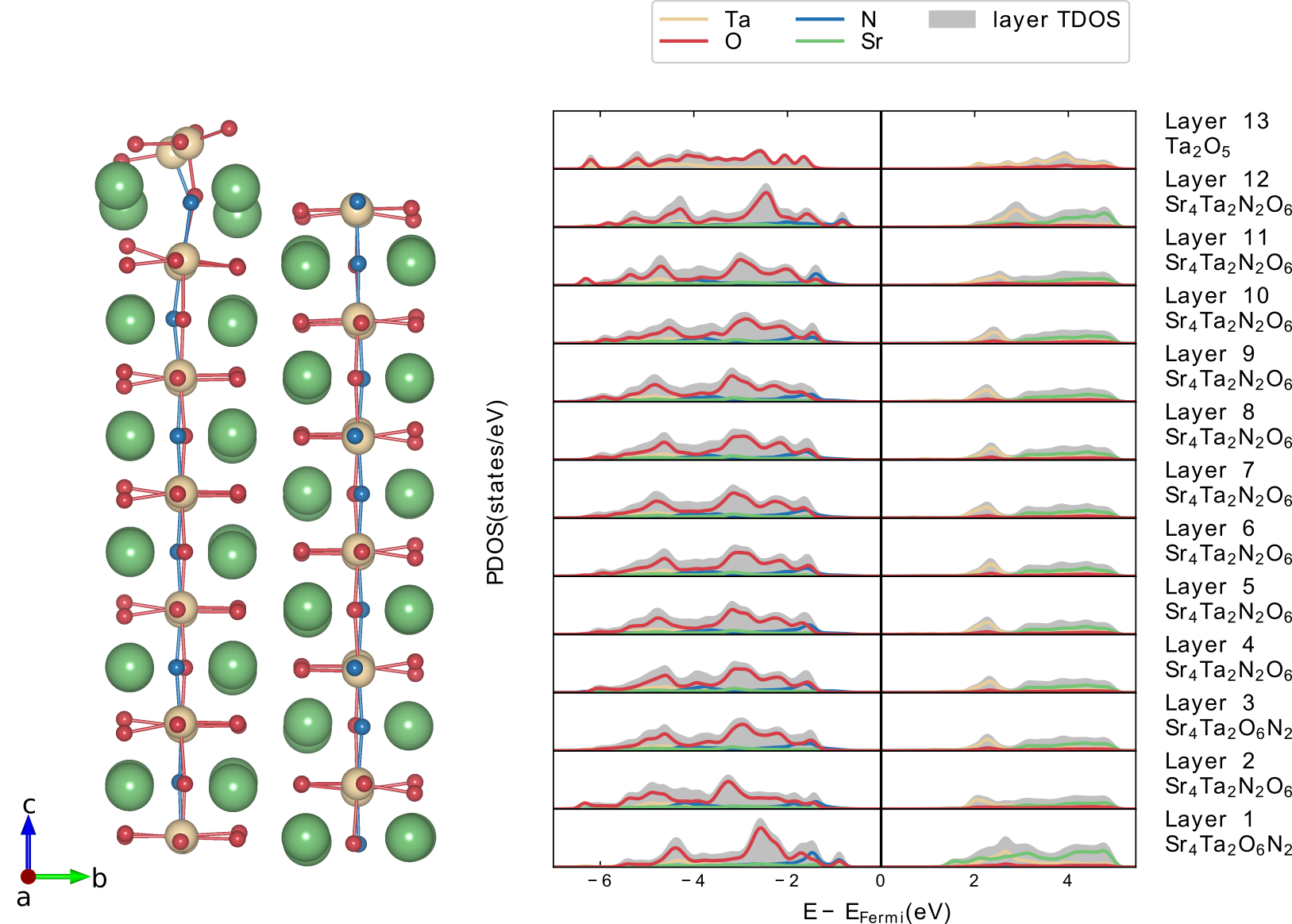}
	\caption{Structure of the (100) stepped surface and its PDOS.}
	\label{fig:100_stepped}
\end{figure}

\begin{figure} [h]
	\centering
	\includegraphics[width=0.5\columnwidth]{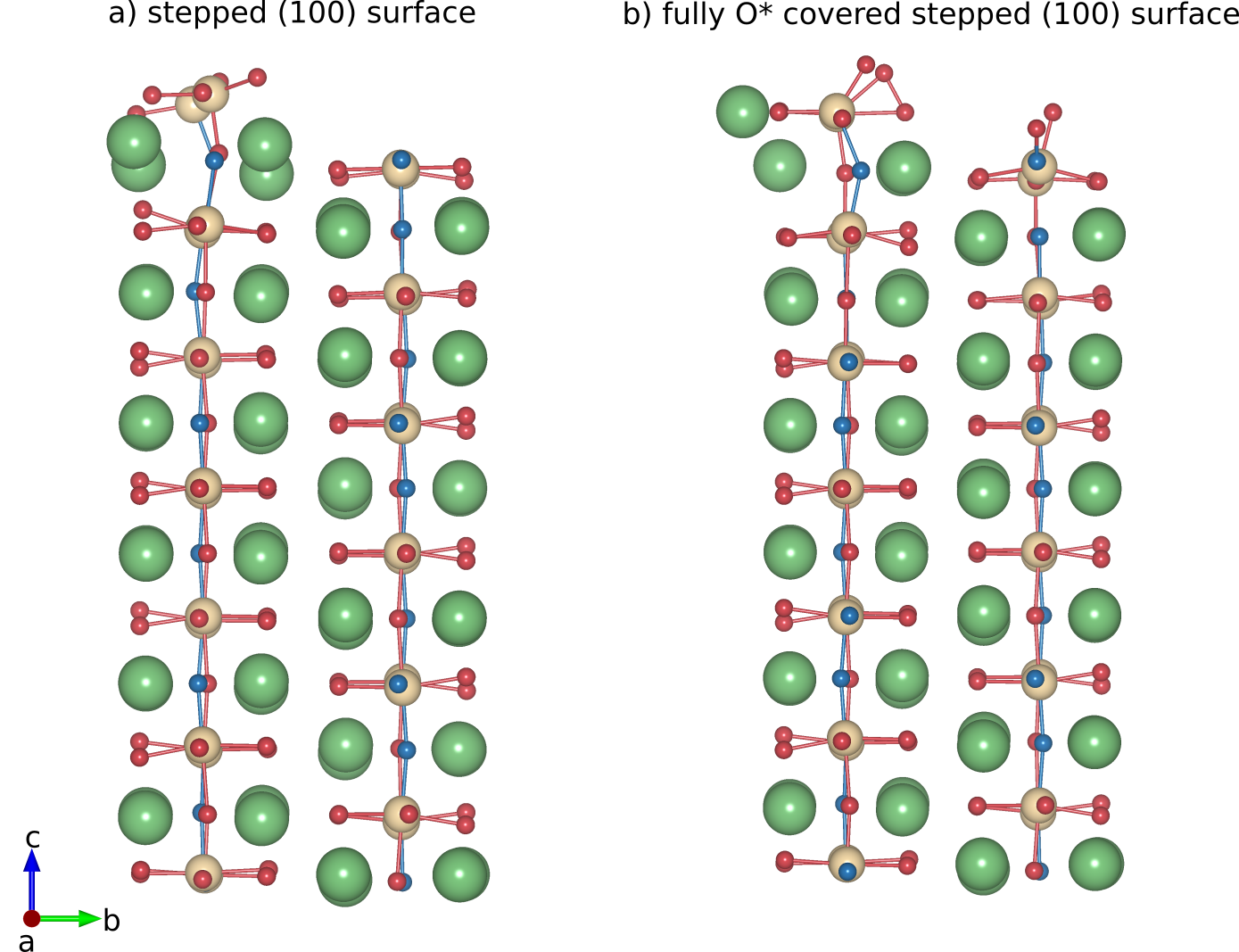}
	\caption{Structure of the a) clean and b) fully O* covered grooved (100) surface.}
	\label{fig:100_step_full_O}
\end{figure}

We hence form a reconstructed (100) slab by removing the Sr atoms belonging to the surface layer of the clean (100) surface (Fig. \ref{fig:100_step_full_O} (a)). The N and O atoms which were coordinated with the Sr atoms are also removed and the surface N atom belonging to the TaON-tetrahedron is substituted with an O coming from the same tetrahedron to cancel the polarity. The reconstruction results in a grooved surface exposing two Ta atoms surrounded by O atoms. The structure and DOS of this surface are shown in Fig. \ref{fig:100_stepped}.

\begin{figure}[h]
	\centering
	\includegraphics[width=0.4\columnwidth]{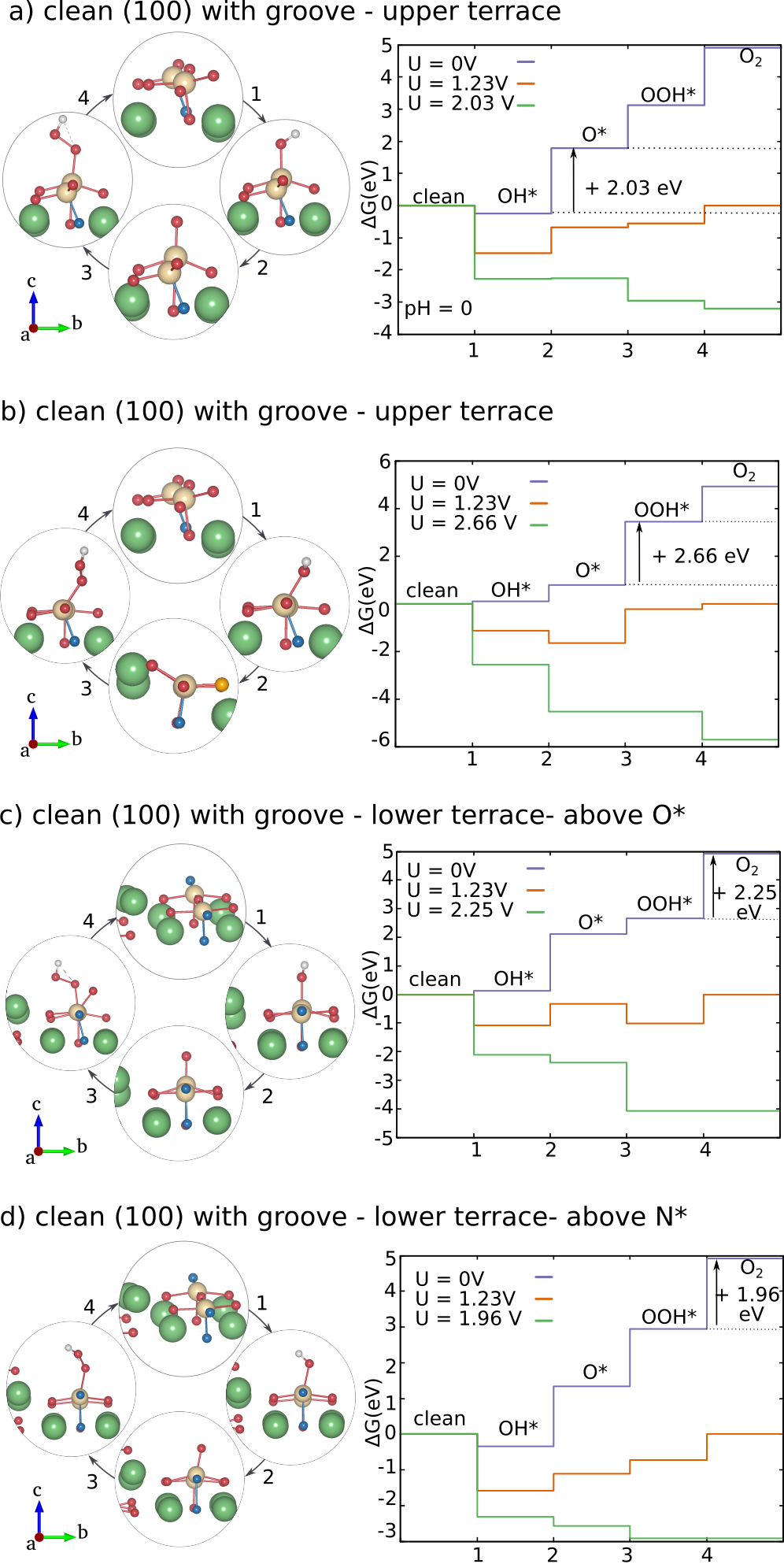}
	\caption{OER steps (left) and Gibbs free energy diagram (right) for the clean grooved (100) surface considering the two sites of a)-b) the upper and c)-d) lower terrace.}
	\label{fig:100_step_overpotentials}
\end{figure}

We investigate the OER on all the sites of the clean grooved (100) surface (Fig. \ref{fig:100_step_overpotentials}). We note that the sites on the upper terrace are 5- and 4-coordinated while the sites of the lower terrace are both 5-coordinated. Moreover, the sites of the upper terrace are surrounded by O atoms in the \textit{xy} plane with the 4-coordinated site also being bonded with a N atom along the \textit{z} direction. The Ta sites on the lower terrace are bonded with N atoms in \textit{xy} plane and one of them also along the \textit{z} direction. We find that the 4-coordinated Ta-site of the upper terrace requires the highest overpotential having as ODS the formation of OOH* (step 3) (Fig. \ref{fig:100_step_overpotentials} b). The Ta site on the lower terrace which is bonded with a N atom in the \textit{z} direction requires the lowest overpotential having as ODS the desorption of the O$_2$ molecule (step 4) (Fig. \ref{fig:100_step_overpotentials} d).  
 
The fully O* covered grooved (100) surface on which we cover all four Ta with O* atoms is shown in (Fig. \ref{fig:100_step_full_O} b). We observe that on the upper terrace where there are no N atoms, the two O are adsorbed in a tilted configuration with one of them making a bond with the surface O atom while the other does not. On the lower terrace, the O* adsorb in a tilted (making a bond with the neighbouring N atom) and in an upright configuration similar to the (001) TaON-terminated surface.

\begin{figure} [h]
	\centering
	\includegraphics[width=0.5\columnwidth]{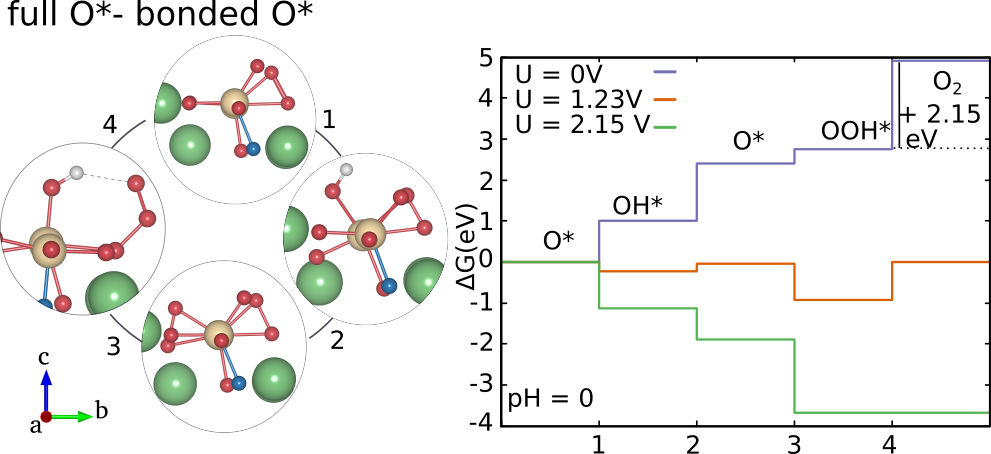}
	\caption{OER steps (left) and Gibbs free energy diagram (right) for the 1-ML-O*-covered grooved (100) surface considering the mechanism on the overcoordinated Ta site.}
	\label{fig:100_step_full_O_mechanism}
\end{figure}

\begin{figure} [h]
	\centering
	\includegraphics[width=0.5\columnwidth]{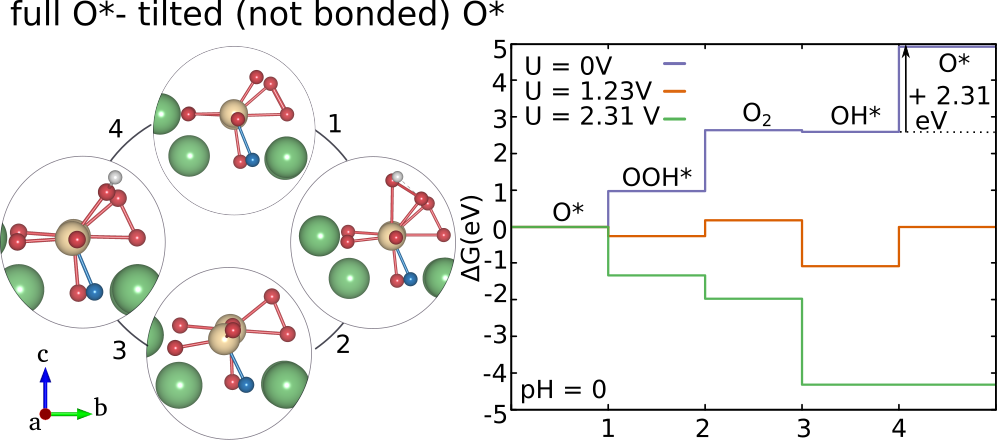}
	\caption{OER steps (left) and Gibbs free energy diagram (right) for the 1-ML-O*-covered grooved (100) surface considering the tilted but not bonded O* as reactive site.}
	\label{fig:100_step_vert_O_up}
\end{figure}

The new mechanism on the tilted and bonded O* of the upper terrace is shown in Fig. \ref{fig:100_step_full_O_mechanism} and we observe a higher overpotential than for the recombination mechanism. The OER on the tilted and non-bonded O* of the upper terrace results in a slightly higher overpotential \ref{fig:100_step_vert_O_up} rendering this site not active for the OER.

\clearpage
\bibliographystyleSI{apsrev4-1}
\bibliographySI{library}

\end{document}